\documentclass[12pt,preprint]{aastex} 
 
\def\pagebreak{\vfill\eject}

\def\Msun{{M$_\odot$}}
\def\sun{\mbox{$_\odot$}}

\def\deg{{$^\circ$}}
\def\half{{\leavevmode\kern.1em\raise.5ex
\hbox{\the\scriptfont0 1}\kern-.1em /
\kern-.15em\lower.25ex\hbox{\the\scriptfont02}}} 
\def\gtsim{\lower.5ex\hbox{$\buildrel > \over\sim$}}
\def\ltsim{\lower.5ex\hbox{$\buildrel < \over\sim$}}
\def\kms{\,km\,s$^{-1}$}
\def\jyb{\,Jy\,beam$^{-1}$}
\def\jybkms{\,Jy\,beam$^{-1}$ km\,s$^{-1}$}
\def\mjybkms{\,mJy\,beam$^{-1}$ km\,s$^{-1}$}
\def\mjyb{\,mJy\,beam$^{-1}$}
\def\co{CO}
\def\cof{CO(J=1$-$0)}
\def\tco{$^{13}$CO}
\def\tcof{$^{13}$CO(J=1$-$0)}
\def\ceo{C$^{18}$O}
\def\ceof{C$^{18}$O(J=1$-$0)}

\def\hco{HCO$^+$}
\def\htco{H$^{13}$CO$^+$}
\def\h{H$_2$}

\def\water{H$_2$O}
\def\methanol{CH$_3$OH}

\def\ra#1#2#3{#1$^{\rm h}$#2$^{\rm m}$#3$^{\rm s}$}
\def\dec#1#2#3{$#1^\circ#2'#3''$}

\def\uchii{UC\,H{\sc ii}}
\def\um{$\mu$m}

\slugcomment{ApJ, accepted}

\begin{document}

\title{Molecular Outflows and a Mid-Infrared Census of the Massive
  Star-Formation Region Associated with IRAS\,18507$+$0121}

\author{
D. S. Shepherd\altaffilmark{1}$^,$\altaffilmark{2}, 
M. S. Povich\altaffilmark{3},
B. A. Whitney\altaffilmark{4},  
T. P. Robitaille\altaffilmark{5},
D. E. A. N\"{u}rnberger\altaffilmark{6},
L. Bronfman\altaffilmark{7},
D. P. Stark\altaffilmark{8}, 
R. Indebetouw\altaffilmark{9},
M. R. Meade\altaffilmark{3}, \& 
B. L. Babler\altaffilmark{3}
}

\vspace{-3mm}
\altaffiltext{1}{National Radio Astronomy Laboratory, P.O. Box O, 1003
  Lopezville Rd, Socorro, NM 87801.}
\altaffiltext{2}{The National Radio Astronomy Observatory is a
  facility of the National Science Foundation operated under
  cooperative agreement by Associated Universities, Inc.}
\altaffiltext{3}{Department of Astronomy, University of Wisconsin at
  Madison, 475 N. Charter St. Madison, WI 53706} 
\altaffiltext{4}{Space Science Institute, 4750 Walnut St. Suite 205,
  Boulder, CO 80301, bwhitney@spacescience.org.}  
\altaffiltext{5}{SUPA, School of Physics and Astronomy, University of
  St Andrews, North Haugh, KY16 9SS, St Andrews, United Kingdom;
  tr9@st-andrews.ac.uk.}  
\altaffiltext{6}{European Southern Observatory, Casilla 19001, 
  Santiago 19, Chile.}
\altaffiltext{7}{Departamento de Astronom\'{\i}a, Universidad de Chile, 
  Casilla 36-D, Santiago, Chile.}
\altaffiltext{8}{Division of Physics, Mathematics \& Astronomy,
  California Institute of Technology, MS 105-24, Pasadena, CA 91125.} 
\altaffiltext{9}{Department of Astronomy, University of Virginia,
  Charlottesville, VA 22903-0818.}

\rightskip=\leftskip 
\vspace{-3mm}

\begin{abstract}
We have observed the central region of the IR-dark cloud filament
associated with IRAS\,18507$+$0121 at millimeter wavelengths in
{\cof}, {\tcof}, and {\ceof} line emission and with the {\it Spitzer
Space Telescope} at mid-IR wavelengths.  Five massive outflows from
two cloud cores were discovered.  Three outflows are centered on or
near an Ultracompact H{\sc ii} (\uchii) region (G34.4+0.23) while the
remaining two outflows originate from the millimeter core
G34.4+0.23\,MM.  Modeling of the SEDs of the mid-IR sources identified
31 young stellar objects in the filament with a combined stellar mass
of $\sim 127 \pm 27$\,\Msun.  An additional 22 sources were identified
as probable cluster members based on the presence of strong 24\,\um\
emission.  The total star formation efficiency in the G34.4 cloud
filament is estimated to be $\sim$\,7\% while the massive and
intermediate mass star formation efficiency in the entire cloud
filament is estimated to be roughly 2\%.  A comparison of the
gravitational binding energy with the outflow kinetic energy suggests
that the compact core containing G34.4+0.23\,MM is being destroyed by its
molecular outflows whereas the outflows associated with more massive
core surrounding the G34.4 \uchii\ region are not likely to totally
disrupt the cloud.  Additionally, a qualitative evaluation of the
region appears to suggest that stars in this region may have formed in
two stages: first lower mass stars formed and then, a few Myrs later,
the more massive stars began to form.

\end{abstract}

\vspace{-.5cm}

\clearpage
\section{INTRODUCTION}

The massive star forming region associated with IRAS\,18507$+$0121
(hereafter IRAS\,18507) is located in an infrared dark cloud that was
mapped by Rathborne et al. (2005) in millimeter and submillimeter
continuum and compared to Spitzer mid-IR emission between 3.5 and
8\um. IRAS\,18507 is roughly $11'$ north of the more famous
ultra-compact (UC) H{\sc ii} region complex G34.26+0.15 (Wood \&
Churchwell 1989; Molinari et al. 1996; Carral \& Welch 1992).  The
location of the infrared dark cloud relative to G34.26+0.15 is shown
in Figure 1.  

Near IRAS\,18507, Miralles et al. (1994) discovered a UC\,H{\sc ii}
region (G34.4+0.23) embedded in a 1000\,M\sun\ molecular cloud traced
by NH$_3$ emission.  The NH$_3$ emission is elongated in the N-S
direction with a total extent of about $7'$, however the $1.5'$
resolution of the observations was not adequate to discern the
structure of the core (Miralles et al. 1994).  Shepherd et al. (2004:
hereafter SNB04) observed IRAS\,18507 in 3\,mm continuum emission,
{\htco} and SiO line emission and in the near-infrared.  They detected
a dense molecular core associated with the UC\,H{\sc ii} region
G34.4+0.23 as well as a dense core and embedded protostellar object
detected only in millimeter continuum emission, G34.4+0.23\,MM
(hereafter G34.4\,MM).  G34.4\,MM appears to be a B2 protostar
surrounded by $\sim 150$ solar masses of warm gas and dust.  Based on
the non-detection of G34.4\,MM at near-IR wavelengths, SNB04 suggested
the source may be in a very early stage of evolution.

Rathborne et al. (2005) identified four clumps in 1.2\,mm continuum
emission which they named MM1-4 (see their figure 1).  MM1 corresponds
to G34.4\,MM, MM2 is coincident with the UC\,H{\sc ii} region
G34.4+0.23, MM3 is about 3$'$ north of G34.4\,MM while MM4 is about
30$''$ south of the UC\,H{\sc ii} region.

IRAS\,18507 was detected in a CS(2-1) survey of IRAS point sources
with far-infrared colors suggestive of UC\,H{\sc ii} regions (Bronfman
et al. 1996).  The source was selected for further high resolution
studies because of its broad line wings, a signature of current star
formation.  By modeling {\hco}, {\htco}, CS and C$^{34}$S spectra
obtained at an angular resolution of $\sim\,16''$ Ramesh et al.\
(1997) demonstrated that the observed line profiles can be explained by
a collapsing hot core of about 800\,M$\sun$ which is hidden behind a
cold ($\sim$\,4\,K) and dense ($3 \times 10^{4}$\,cm$^{-3}$) envelope
of about 200\,M$\sun$.  

The kinematic distance to IRAS\,18507 has been reported to be between
3.7 and 3.9\,kpc based on molecular line velocities that vary from 55
to 59{\kms} depending on the line observed and the position within the
cloud (3.9\,kpc: Molinari et al. 1996, Shepherd et al. 2004, Zhang et
al. 2005; 3.8\,kpc: Carral \& Welch 1992, Miralles et al. 1994, Ramesh
et al. 1997; \& 3.7\,kpc: Rathborne et al. 2005).  Kinematic distances
can easily have an error of more than 5 or even 10\% due to observed
deviations from circular symmetry and circular motion in our galaxy.
Thus, a difference of 0.2\,kpc is easily within the errors for this
type of distance determination.  Here we adopt the distance of
3.9\,kpc to be consistent with Molinari et al. (1996, 1998), SNB04 and
Zhang et al. (2005).  If the closer distance of 3.7\,kpc is more
appropriate, then mass estimates derived in this work should be
decreased by about 10\% since the mass is proportional to $D^2$.

Fa\'{u}ndez et al (2004) mapped the IR-dark cloud with the SEST
telescope in 1.2\,mm continuum emission and derived mass of
2000\,M$\sun$ in the main condensation based on a dust temperature of
28\deg K.  The total mass of the filament was estimated to be about
3700\,M\sun\ where 1000\,M\sun\ is associated with G34.4\,MM and
2700\,M\sun\ is associated with the G34.4 {\uchii} region (Bronfman,
personal communication).  Rathborne et al (2005) mapped the dark cloud
with the IRAM\,30 telescope in 1.2\,mm continuum emission and derived
a cloud mass of $\sim 7500$\,M\sun\ based on a dust temperature of
34\deg K, two times more mass than derived by Fa\'{u}ndez et al.  In
this work, we will assume the temperature and mass estimates of
Rathborne et al. since a dust temperature of 34\deg K would appear to
be more representative of massive star forming regions.  However, a
clear reason for why the derived temperature and, hence, mass
estimates differ by such a large amount is not clear.

The IRAS\,18507 region is also associated with variable {\water},
{\methanol} and OH maser emission (Scalise et al. 1989; Palla et
al. 1991; Schutte et al. 1993; Miralles et al. 1994; Szymczak et
al. 2000; Wang et al. 2006; Edris et al. 2007).  Molinari et
al. (1996, 1998) observed IRAS\,18507 (labeled Mol74 in their papers)
and estimated a deconvolved size of the UC\,H{\sc ii} region of 0.7''
(0.013\,pc at D=3.9\,kpc).

Based on observations with the NRAO 12\,m telescope, Zhang et
al. (2005) report that IRAS\,18507 is not associated with molecular
outflows.  The lack of outflow activity contrasts with the presence of
multiple star formation tracers (masers, dense gas, HII regions and
millimeter continuum).

In this work, we present observations of the dense molecular gas near
IRAS\,18507 at millimeter wavelengths in {\cof}, {\tcof}, and {\ceof}
line emission to map the dense gas morphology and determine if
there are massive outflows associated with IRAS\,18507.  Archive data
from the {\it Spitzer Space Telescope} GLIMPSE survey at mid-IR
wavelengths were studied to obtain a census of the massive and
intermediate-mass star cluster properties over the entire
infrared-dark cloud filament.  Section 2 gives an overview of the
observations, section 3 presents the results and section 4 discusses
implications.

\section{OBSERVATIONS}

\subsection{Observations in the 3~mm band}

Simultaneous observations in {\cof}, {\tcof} and {\ceof} lines were
made with the Owens Valley Radio Observatory (OVRO) array of six
10.4~m telescopes on 2004 March 27 \& 29 and 2004 April 16 \& 18.
Projected baselines ranging from about 12 to 115 meters provided
sensitivity to structures up to about $20''$ with $\sim 5''$
resolution.
Two fields were observed as a Nyquist-sampled mosaic.  The
observations were alternated between fields every 9 minutes to assure
that the $uv$ coverage in each field was similar.  The total
integration time on each field was approximately 8.8 hours.
Cryogenically cooled SIS receivers operating at 4~K produced typical
single sideband system temperatures of about 400~K.  The gain
calibrator was the quasar $1751+096$ and the passband calibrators were
3C\,345, 3C\,454.3 and 3C\,273.  Observations of Uranus provided the
flux density calibration scale with an estimated uncertainty of $\sim
15$\%.  Calibration was carried out using the Caltech MMA data
reduction package (Scoville et al. 1993).  Images were produced using
the MIRIAD software package (Sault, Teuben, \& Wright 1995) and
deconvolved with a Maximum Entropy (MEM) algorithm.  Data analysis was
performed using both the MIRIAD package and the CASA (Common Astronomy
Software Applications) package.

The spectral band pass for each line was centered on a velocity of
$57.0$\kms.  The {\co} images have a synthesized beam of $3.83''
\times 3.45''$ (FWHM) at P.A. $-59.7$\deg, spectral resolution
1.3{\kms} and RMS noise 60\mjyb.
The {\tco} images have a synthesized beam of $4.04'' \times 3.83''$
(FWHM) at P.A. $-49.9$\deg, spectral resolution 1.36{\kms} and RMS
noise 44.0\mjyb.
The {\ceo} images have a synthesized beam of $4.04'' \times 3.83''$
(FWHM) at P.A. $-50.1$\deg, spectral resolution 0.683{\kms} and RMS
noise 56.0\mjyb.

\subsection{Spitzer images}

Mid-infrared images at 3.6, 4.6, 5.8 and 8.0 \um\ were obtained using
the Infrared Array Camera (IRAC; Fazio et al.\ 2004) of the {\it
Spitzer Space Telescope} as part of the Galactic Legacy Infrared
Mid-Plane Survey Extraordinaire (GLIMPSE; Benjamin et al.\ 2003).
Image mosaics were created by the GLIMPSE pipeline after image
artifacts such as cosmic rays, stray light, column pull-down, and
banding were removed.\footnote{Details of the data processing can be
found at http://www.astro.wisc.edu/glimpse/docs.html}.  Point sources
were extracted from the GLIMPSE images using a modified version of
Daophot (Stetson 1987). The astrometric uncertainty for point sources
is $\sim 0.3''$.  The GLIMPSE $5\sigma$ point-source sensitivities are
approximately 0.3, 0.3, 1.2, and 0.7 mJy in the four IRAC band passes,
respectively.  Point source sensitivities were generally higher in
areas of bright diffuse background emission.  The fluxes of point
sources that are detected at $\geq 5\sigma$ twice in one IRAC band or
once in each of two bands are entered in the GLIMPSE Point Source
Archive.  Each Archive entry includes the corresponding $J$, $H$, and
$K_s$ fluxes from the 2MASS Point Source Catalog (Skrutskie et al.\
2006) if that source was detected in 2MASS.

The Multi-band Imaging Photometer for {\it Spitzer} (MIPS; Rieke et
al.\ 2004) was used to image the entire GLIMPSE survey area at 24 and
70 \um\ as part of the MIPSGAL survey (Carey et al. 2005).  
Fluxes at 24\,\um\ for point sources within the G34.4 IR dark cloud
were extracted from the frame using the GLIMPSE point-source
extractor.  The relative uncertainty between two sources in a MIPSGAL
image is roughly 0.3 to 0.5 pixels while the absolute positional
uncertainty is $\sim 1''$.

\section{RESULTS}

Figure 1 presents a four-color (3.6, 4.6, 5.8, \& 8.0 \um) GLIMPSE
image of the G34.4 environment in Galactic coordinates (North points
to the upper left corner).  The ``cometary'' \uchii\ region
G34.26+0.15 is located on the right side of the image.  An IR-bright
nebula extends below the G34.26 UC H{\sc ii} region. The UC H{\sc ii}
region contains a cluster of OB stars, the most luminous of which is
an O6.5 spectral type (Wood \& Churchwell 1989).  The nebula is bright
in the IRAC bands primarily due to PAH emission features stimulated
by UV radiation from the cluster.  The subject of this paper is the
dark filament in the upper left which is forming early B stars.  The
mid-IR emission from the UC\,H{\sc ii} region G34.4+0.23 is
significantly less than that of G34.26, as expected due to the lower
luminosity of the central source.  Figure 2 presents a close-up of the
IR-dark filament, now displayed in Equatorial coordinates with North
pointing upwards.  Along the dark filament, North and South of G34.4,
are embedded sources seen predominantly by their 4.5\,$\mu$m (green)
emission.  Our OVRO CO mosaic is centered on the southern core near
the UC\,H{\sc ii} region G34.4 and the millimeter core G34.4\,MM
located $\sim 40''$ north of the UC\,H{\sc ii} region.

Figure 3 presents the red and blue-shifted emission near G34.4\,MM and
G34.4 along with a 2\,$\mu$m K band image from SNB04.  Neither the
millimeter core nor the UC\,H{\sc ii} region is visible in the K band
image.  Both regions show complex high-velocity structure suggesting
the presence of multiple outflows.

Figure 4 shows the integrated emission (moment 0) in {\ceo} and {\tco}
along with the red and blue-shifted {\tco} emission.  The bottom two
panels in figure 4 show moment 1 maps of {\ceo} \& {\tco} in which the
brightness of the color is proportional to the intensity in the moment
0 image.  This type of image provides some insight into how the dense
cloud structure and kinematics are being affected by the molecular
outflows.

Figures 5, 6 \& 7 present channels maps of CO, {\tco} and {\ceo}
emission, respectively.

\subsection{Molecular outflows}

At least two outflows can be seen emanating from G34.4\,MM (labeled A
and B in the upper left panel of Fig. 3) while at least three
outflows originate from the southern core near the G34.4 UC\,H{\sc ii}
region (labeled C, D and E).

The velocity structure of the outflows from the G34.4\,MM region is
consistent with a single outflow dominating the red-shifted kinematics
(outflow G34.4:A).  There is a weak blue-shifted counter flow
extending to the NE that can be traced to G34.4\,MM.  The red-shifted
lobe of outflow G34.4:A appears to have little or no associated
blue-shifted emission suggesting that it is oriented outside of the
plane of the sky (the blue-shifted emission in the red lobe appears to
be from the collimated flow G34.4:E coming from the southern region).
The full-opening angle of the red-shifted lobe is $\sim 43$\deg.

Marginal evidence for an outflow perpendicular to outflow G34.4:A is
seen in the red and blue-shifted emission clumps along the axis
labeled 'B'.  No source has been detected at millimeter wavelengths
which could drive outflow G34.4:B.  The source could be embedded in
G34.4\,MM and not resolved at the resolution of $\sim 5''$ of SNB04.  

The southern molecular core containing the G34.4 UC\,H{\sc ii} region
contains three overlapping flows seen in the CO emission (flows 'C',
'D' and 'E').  Blue-shifted CO, {\tco} and {\ceo} gas is seen toward
the G34.4 UC\,H{\sc ii} region extending to the West (flow G34.4:C).
This outflow is seen most clearly in {\tco} and {\ceo} emission in
Figure 4.  {\tco} blue-shifted emission is traced back to the H{\sc
ii} region (top right panel in Fig. 4) while the bottom two panels
illustrating the velocity structure in the dense gas show that
blue-shifted emission extends from the UC\,H{\sc ii} region to the
west.  A possible red-shifted outflow lobe east of the UC\,H{\sc ii}
may be visible in {\tco} emission in the top right panel in Fig. 4
although the identification is not definitive.

The NE-SW flow (G34.4:D) appears to be centered about $10''$ SE of the
UC\,H{\sc ii} region, just off the peak of a dense {\ceo} clump.  The
flow appears to be relatively symmetric in the red and blue-shifted
gas distribution although the actual morphology is considerably
confused due to overlapping outflows.  If the flow is symmetric, then
one would expect the driving source to be located near the small,
filled triangle at \ra{18}{53}{19.0}, \dec{01}{24}{35} along the
G34.4:D outflow axis in Figs. 3 \& 4.   

The SE-NW flow (G34.4:E) is seen clearly in both CO and {\tco}
emission.  If the driving source is located between the red- and
blue-shifted emission along the outflow axis, it would be centered
about $5''$ S-SW of the UC\,H{\sc ii} region (indicated by a small,
filled triangle at \ra{18}{53}{18.6}, \dec{01}{24}{43} along the
G34.4:E outflow axis in Figs. 3 \& 4).  The outflow is not symmetric:
the red-shifted emission lobe extends about $20''$ to the SE while the
blue-shifted lobe extends roughly $60''$ to the NW, overlapping the
blue-shifted lobe from the G34.4 UC\,H{\sc ii} region flow and the
red-shifted lobe of flow G34.4:A farther north.

Outflow G34.4:A has a position angle of +58{\deg} (measured from North to
the axis of the blue-shifted lobe) while G34.4:B P.A. = --38{\deg} and
the difference in orientation between outflow axes is 96{\deg}.
The positions angles of the three outflows near the G34.4 UC\,H{\sc ii}
region are: 
G34.4:C P.A. = --100{\deg}; 
G34.4:D P.A. =   +38{\deg}; 
G34.4:E P.A. =  --34{\deg} 
and the maximum difference in orientation between the axes is
138{\deg}. 

\subsection{Mass \& Kinematics of the Outflows} 

The mass associated with CO line emission is calculated following
Scoville et al. (1986).  The CO excitation temperature is taken to be
34 K which is the dust temperature in MM1 derived from fits to the SED
(Rathborne et al. 2005).  We assume the gas is in LTE with
[CO]/[\h]~=~$10^{-4}$, [CO]/[\tco]~=~51 and [CO]/[{\ceo}] = 378 at
the galacto-centric distance of 5.8~kpc (Wilson \& Rood 1994).  

The CO optical depth at red and blue-shifted velocities is estimated
from the CO/{\tco} ratio of spectra taken at four positions in the
mosaic image that have been convolved with a $10''$ beam.  The open
triangles in the top, center panel of Fig. 4 show the locations where
the spectra were taken, while Figure 8 shows the spectra taken at those
positions.  In channels where no {\tco} is detected, we assume that
the CO is optically thin.  The CO and {\tco} line profiles show a dip
near $v_{\rm LSR}$ indicating that the lines suffer from
self-absorption and/or the interferometer is missing extended flux at
lower velocities.

Because multiple, overlapping flows are present, it is not possible to
obtain a reasonable estimate of the inclination of each flow.  Thus,
we assume an inclination angle of 45\deg\ which minimizes errors
introduced by inclination effects for flows that are not near the
plane of the sky or along the line-of-sight.  Table 1 summarizes the
physical properties of the molecular gas in the combined outflows
originating from G34.4\,MM and those centered near the UC\,H{\sc ii}
region.  The total flow mass $M_f$ is given by $\sum M_i$ where $M_i$
is the flow mass in velocity channel $i$ corrected for optical depth.
The momentum $P$ is given by $\sum M_i v_i$ and the kinetic energy E
by $\frac{1}{2} \sum M_i v_i^2$ where $v_i$ is the central velocity of
the channel relative to $v_{LSR}$.

The dynamical timescale $t_d$ is calculated using $R_{f}/<V>$, where
the intensity-weighted velocity $<V>$ is given by $P/(\sum M_i)$
(Cabrit \& Bertout 1990) and $R_{f}$ is the average outflow radius of
the flows.  This method of calculating $t_d$ is really only correct
for a single outflow; however, an independent outflow age cannot be
determined for both flows in the cores since they cannot be separated
near the central position.  Because of this uncertainty, the flow
dynamical timescale could easily be in error by a factor of 2 to 5.

The mass outflow rate $\dot M_{f}$ is $\sum M_i/t_d$, the force F is
$P/t_d$ and the mechanical luminosity L is $E/t_d$.  Since these
values are dependent on $t_d^{-1}$ uncertainties in $t_d$ are directly
applicable to $\dot M_{f}$, F, and L.

The combined outflows from the G34.4 core have a total mass of
111~M\sun\ and energy, E $= 7.8 \times 10^{46}$~ergs.  The G34.4\,MM
outflows have a combined mass $34.8$~M\sun, E $= 4.9 \times
10^{46}$~ergs.  Despite the uncertainties, the flow masses and
energies are consistent with those for outflows driven by young, early
B stars.

\subsection{The dense gas cores}

The {\ceo} and {\tco} moment 0 images (Fig. 4) show clearly that the
two cores have very different characteristics.  The southern core
centered on the G34.4 UC\,H{\sc ii} region is extended, with a largest
spatial scale of about 1.4\,pc at a distance of 3.9\,kpc.  The
northern core, centered on G34.4\,MM, is significantly more compact
with higher peak emission than the southern cloud.  The diameter of
the G34.4\,MM core is about 0.2\,pc and is surrounded by faint,
diffuse {\tco} emission that extends out to about 1\,pc in size.

The total mass of the 2 cores as measured from {\ceo} is given in
Table 2. {\ceo} is assumed to be optically thin although this is not
likely to be true near $v_{LSR}$ at the peak of the line profile.  The
extended, southern cloud in which the UC\,H{\sc ii} region is embedded
is almost 10 times more massive (692\,M{\sun}) than the G34.4\,MM core
(75\,M{\sun}).

As Fig. 8 illustrates, the {\ceo} line appears to be suffering from
missing flux due to the lack of short baselines.  This is most
apparent in the spectra taken SE of the UC\,H{\sc ii} region position
where {\ceo} is more extended and there is a clear double peaked line
profile.  The {\ceo} line is single peaked closer to the cloud core
suggesting that the interferometer is recovering most of the flux
density and we are not seeing significant self-absorption.  Thus, the
mass estimates based on {\ceo} are lower limits.

The G34.4 and G34.4\,MM molecular cores have a projected separation of
about $40''$ (0.75\,pc at 3.9\,kpc).  Their LSR velocities
as measured by the peak of the {\ceo} line differs by 1.4{\kms}.

\subsection{Mid-Infrared Sources: GLIMPSE \& MIPSGAL}

G34.4\,MM and the G34.4 UC\,H{\sc ii} region are embedded in an
extended, filamentary IR-dark cloud complex seen against the diffuse
PAH background emission which is visible primarily in the Spitzer
8.0\,$\mu$m band (Figures 1 \& 2).  The G34.4\,MM core is associated
with diffuse emission in all 4 IRAC bands, but is particularly
noticeable at 4.5 \um. The [4.5] band contains the H {\sc i}
Br$\alpha$ line along with some molecular forbidden lines, notably the
CO 1-0 P(8) fundamental band-head, but, unlike the other 3 bands, does
not contain any PAH emission feature.  The [4.5] band also contains
several \h\ lines (e.g. the \h\ 0-0 S(9) line at 4.694 \um) which
trace shocks in some massive outflows (e.g. DR21: Smith et al.  2006;
W75N: Davis et al. 2006).  Thus, the [4.5] band appears to readily
trace ionized and/or shocked gas.

Figure 9 shows the IRAC 4.5\,$\mu$m image in greyscale with {\tco}
contours overlayed for reference.  The plus symbols represent the
locations of the \uchii\ region and G34.4\,MM core.  The open diamonds
just south and south-east of the \uchii\ region represent the
predicted locations of sources driving outflows G34.4:E
(\ra{18}{53}{18.6}, \dec{01}{24}{43}) and G34.4:D (\ra{18}{53}{19.0},
\dec{01}{24}{35}) based on CO and {\tco} emission.  While the
predicted position of the YSO that was identified based on the CO
outflow morphology should be considered no better than an educated
guess, it is interesting that each position is associated with a
mid-IR source to within a few arcsec.  The characteristics of these
sources along with other sources in the cluster are discussed below.

The spectral energy distributions (SEDs) were analyzed by fitting
model SEDs to each observed source from a large pre-computed grid of
YSO models (Robitaille et al. 2006, 2007).  The grid consists of 20,000
2-D YSO radiation transfer models spanning a complete range of stellar
mass and evolutionary stage.  Each YSO model outputs SEDs at ten
viewing angles (inclinations), so the fitter actually has 200,000 SEDs
to choose from.  The YSO models also output SEDs for each of 50
aperture sizes.  Thus, if the distance to the star formation region is
known, only the model SEDs with the appropriate aperture size compared
to the observed photometry are used.  A grid of stellar atmospheres
were also incorporated into the fitting algorithm in order to
facilitate the separation of main-sequence and evolved stars from YSOs
(Brott \& Hauschildt 2005; Kurucz 1993).  The fitting algorithm
includes a foreground extinction component to the model which allows
one to easily distinguish highly reddened stars from YSOs which have
an additional emission component in the mid-infrared.

The 24\,\um\ fluxes were cross correlated with positions of the 2199
GLIMPSE Archive sources lying within a 3.9\arcmin\ radius of the G34.4
\uchii\ region to account for observational errors in addition to
photon counting statistics.  For the purposes of the fitting, all flux
uncertainties that fell below the 10\% level were reset to a minimum
value of 10\%.  The SED fitter was then run on this combined 8-band
source list, fitting only the fluxes of sources detected in 4 or more
bands.  Sources with SEDs well fit by a stellar atmosphere are not
likely to be cluster members.  Sources with a $\chi^2$ per flux data
point ($\chi^2/N_{\rm data}$) of 4.8 provided a reasonable dividing
line between sources that, by eye, appeared to be well-fit by stellar
atmospheres and those that were not well-fit due to IR excesses.  Of
the total 2199 sources in the GLIMPSE Archive in this region, 868 were
detected in four or more bands.  Of those, 829 sources were well-fit
by stellar atmospheres and thus are likely foreground or background
stars rather than YSO cluster members.  The remaining sources were fit
with YSO models and 27 of these met the $\chi^2/N_{\rm data}\leq 4.8$
criterion. Of the 9 sources that were not well-fit either by stellar
atmospheres or YSO SEDs, 1 appears to be a YSO and 8 are probably Archive
sources with improperly extracted fluxes in one or more bands or
there was a mismatch of GLIMPSE with 2MASS or MIPS sources.

An overview of the results of the SED fitting procedure is presented
in Figure 10 and Tables 3 \& 4.  Most of the filamentary dark cloud is
visible in the three-color image (Fig. 10) in which red corresponds to
MIPS 24\,$\mu$m, green to GLIMPSE 8\,$\mu$m and blue to GLIMPSE
4.5\,$\mu$m. The G34.4 complex is located near the center of Fig. 10
(in the region outlined with a large, white circle).  Point sources
are marked by small circles color-coded by source type as identified
by the Fit flag in Table 3.  Sources that were well-fit as YSOs are
marked by green circles and given a Fit flag of 1 in Table 3.  The
yellow circle identifies a probable YSO that was not well-fit (Fit =
2).  Blue circles identify sources that were well-fit by stellar
photospheres.  MIPS 24 \um\ point sources that are detected in fewer
than 3 of the 2MASS+IRAC bands and hence were not run through the
fitter (Fit=4) are identified by red circles; the majority of these
are probably YSOs.  The cyan circle (source 32 located between sources
8 and 10 in Fig. 10) marks the Archive source that coincides with the
proposed origin of outflow G34.4:D, it is detected only at 3.6 and
4.5\,\um, and hence was not run through the fitter (Fit=3).  The area
covered by the OVRO observations is contained primarily within the
large white circle, which is referred to as ``region C'' in Table 3.
``Region N'' is identified by the box north of region C, while
``region S'' is identified by the box to the south.  The G34.4
UC\,H{\sc II} region appears white in this image because it is bright
in diffuse emission at all wavelengths.  There are two prominent MIPS
sources: G34.4\,MM and another, fainter source located about 2\arcmin\
north of G34.4\,MM at position \ra{18}{53}{20.602},
\dec{+01}{28}{25.61}.

Table 3 presents a summary of the positions and fluxes in each band
for all identified probable YSO cluster members in and near the dark
cloud.  Table 4 presents the best fit model parameters for each YSO
with enough flux points to be fitted.  Output model parameters are:
interstellar extinction to the source, $A_V$; stellar mass, $M_\star$;
total luminosity of the YSO, $L_\star$; envelope accretion rate,
$\dot{M}_{\rm env}$; and inclination, $i$.  Note that $A_V$ does not
include the extinction produced by a circumstellar disk or envelope
and thus it represents a lower limit for YSOs with significant
circumstellar material.  In general, each source can be well fit by
multiple YSO SED models and this generates a corresponding range of
the best-fit model parameters. For well-fit YSOs, results of all fits
with $\Delta \chi^2/N_{\rm data} \leq 1$ relative to the best-fit
model are reported. E.g., for a typical source detected in $N_{\rm
data}=5$ bands, all fits with $\Delta\chi^2 \geq 5$ with a formal
probability relative to the best fit of less than
$\exp(-\Delta\chi^2/2)=0.08$ are excluded. The cumulative probability
distribution for each parameter is then constructed using the
Gaussian-weighted probability of each fit relative to the best fit.  A
range of parameter values that has a greater than 95\% probability of
containing the actual value for the source is then determined and
reported in Table 4.
  
\noindent {\bf The central region of the filament:} \\
In the region for which CO outflow data is determined, 7 YSOs with
good fits have been identified (sources 6-12), along with one
poorly-fit, possibly massive YSO (source 31 in tables 3 \& 4) located
on the northeast edge of the G34.4\,MM 24\,\um\ emission.  Source 12,
the highly luminous, the mid-IR counterpart to G34.4\,MM, is most
likely the driving source of outflow G34.4:A.  Marked by a larger
green circle in Figure 10, source 12 is partially resolved in the
GLIMPSE images (this is seen most clearly in the 4.5\,\um\ greyscale
image in Fig. 9).  The morphology of source 12/G34.4\,MM is consistent
with two dust emission lobes separated by an absorption lane due to a
disk where the lobes are in rough alignment with the axis of outflow
G34.4:A.

The fluxes in the two lobes of source 12/G34.4\,MM in each GLIMPSE
band was summed and combined with the point-source fluxes from {\it
MSX} 21.3 \um\ (Price et al.\ 2001), MIPS 24 \um\, and MIPS 70\um\ to
model the source SED across the mid-IR.  The 10 best fits to source
12/G34.4\,MM are plotted in Figure 11.  The fit parameters are
well-constrained and the best fit has $\chi^2=6.62$ which is
significantly lower than the second-best fit with $\chi^2=20.79$.  The
best fit parameters suggest a massive YSO with $M_\star \sim
$14\,M{\sun}, $L_\star \sim $ 9,400\,L{\sun}. The source is deeply
embedded, behind greater than $A_{V}\sim 40$ mag of extinction and it
is young with a high envelope accretion rate of $\dot{M}_{\rm env}\sim
10^{-3}$ \Msun\ yr$^{-1}$.

The inclination angle of the best-fit model is $i=56${\deg},
consistent with the orientation of outflow G34.4:A which has a wide
opening angle of 43{\deg} but does not intersect the plane of the sky.
This interpretation assumes that a single YSO dominates the flux
density and geometry of the mid-IR emission.  Given that two outflows
have been identified emanating from the G34.4\,MM core and the far-IR
fluxes exceed those predicted by a single YSO SED, more than one YSO
is likely present.  However, the resolution of the GLIMPSE images is
not adequate to clearly distinguish between sources that are likely to
drive outflows G34.4:A and G34.4:B.

For the remaining six well-fit YSOs in region C, the best fit models
are presented in Figure 12a while the distribution of their
corresponding best-fit masses and luminosities are shown in Figure
12b.  Of these, two YSOs (sources 10 \& 11) are nearly coincident with
the \uchii\ region.

Source 11 is a luminous YSO located near the projected center of the
{\uchii} region and appears to be the most likely candidate to drive
outflow G34.4:C.  The GLIMPSE position differs from the 6\,cm ionized
gas peak by $\sim 1.3''$.  This could be due to differences in
astrometry between the radio and IR data; however, the discrepancy
exceeds the GLIMPSE astrometric uncertainty of 0.3\arcsec\ and the Very
Large Array estimated uncertainty of $\sim 0.1''$.  It is more likely
that the centimeter peak tracing the ionized gas is offset from the
actual YSO.  This could easily happen if the ionized gas morphology
closely follows the confines of a bipolar outflow cavity.  It is also
possible that the center of the source is not detected in the IRAC
bands due to high extinction or the bright diffuse background of the
\uchii\ region itself. We may only see the warm outflow cavity
associated with the blue-shifted lobe in the mid-IR that is displaced
from the centimeter continuum emission.

Both sources 10 and 11 suffer from confusion with the \uchii\ region
at IR wavelengths longward of the IRAC bands.  The \uchii\ region
emits very strongly in all 4 {\it MSX} bands and in MIPS 24 and
70\,\um.  These fluxes were used as upper limits for the fits to
sources 10 and 11, but they are very high limits (see Figure 12a), and
the fits to both sources are poorly constrained.  The best fit results
for source 11 suggest $M_\star = $2--10\,M{\sun} and $L_\star =
$200--5,200\,L{\sun}, but the probability distribution of fit
parameters shown in Figure 12b is skewed and allows for higher masses
and luminosities up to $M_\star\sim 20$ \Msun, $L_\star\sim 45,000$
L{\sun} (albeit at a formal probability level of only a few percent).
SNB04 determined a spectral type of B0.5 ($L_\star \sim
40,000$\,L{\sun}) for the central star of the G34.4 \uchii\ region
assuming a single zero-age main sequence (ZAMS) star is producing the
observed Lyman continuum flux.  Thus, if source 10 is the ionizing
star of the \uchii\ region, the higher values of mass and luminosity
are more plausible.

Due to its physical location between the red and blue-shifted lobes of
outflow G34.4:E, source 10 appears to be the most probable driver of
this outflow.  The best fit results suggest that it is a YSO with
M$_\star = $0.5--10\,M{\sun} and L$_\star = $45--4,500\,L{\sun}.
Again, the association of this YSO with strong outflow activity
suggests that the upper end of this range of mass and luminosity may
be more appropriate.

Source 32 (cyan circle), the potential driver of outflow G34.4:D, is
detected in only two IRAC bands ([3.6] and [4.5]) at the $5\sigma$
level sufficient for inclusion into the GLIMPSE Archive.
Based on these two bands, source 32 has a red color of
$[3.6]-[4.5]=1.3$ which is consistent with it being a YSO (Robitaille
et al. 2006).  The lack of GLIMPSE Archive fluxes in bands [5.8] and
[8.0] means that this source was not detected at the $3\sigma$ level
in those two bands, corresponding to magnitudes fainter than 13.0 in
each band\footnote{See the GLIMPSE Quality Assurance Document at
http://www.astro.wisc.edu/glimpse/GQA-master.pdf}.  The colors of YSO
models for very young sources predict $[3.6]-[5.8]>0.4$ and
$[5.8]-[8.0]>0.4$, corresponding to magnitudes fainter than 13.6 and
13.2 in bands [5.8] and [8.0], respectively (Robitaille et al.  2006).
Thus the non-detection in bands [5.8] and [8.0] is consistent with the
interpretation that source 32 is a YSO.

Source 32 is also not detected in the MIPS 24 \um\ band.  However,
given the proximity to the very bright \uchii\ and lower resolution of
MIPS compared to IRAC, this source is confused with the PSF wings of
the \uchii\ region.  Based on an estimate of the MIPS sensitivity in a
confused region, a $3 \sigma$ upper limit of 50 mJy (magnitude 5.4) is
expected for the 24 \um\ flux.  This could potentially allow a very
red $[3.6]-[24]\approx 8.6$ color for this source.  We therefore
conclude that the mid-IR colors of source 32 are consistent with those
for a YSO, however the strongest evidence for this remains the fact
that it lies along the CO outflow axis of G34.4:D near where one might
expect the driving source to be located.

In summary, 9 YSOs have been identified and characterized in region C near
G34.4\,MM and the {\uchii} region.  Four or five are probably massive
with $M_\star > 8$\,M{\sun} while the others are low-to-intermediate
mass with $M_\star$ between 0.5 and 8\,M{\sun}.

\noindent {\bf The northern filament:} \\
North of the G34.4\,MM and {\uchii} region complex (in region N) 17
YSOs have been successfully identified based on fits to the SEDs.
Most of these YSOs closely trace the IR-dark cloud structure.  At the
extreme north end of the field there is another 24\,$\mu$m bright
source (source 29 in Tables 3 \& 4) that may be similar to but less
massive than G34.4\,MM.  
The SED is well-fit with $\chi^2=17.39$ suggesting a YSO with M$_\star
\sim $4.8--6.8\,M{\sun} and L$_\star \sim $170--340\,L{\sun}.  The
remaining 16 YSOs in region N are less massive than Source 29 with
mass estimates between about 0.5 and 6.5\,M{\sun}.

\noindent {\bf The southern filament:} \\ 
The southern part of the IR-dark filament (region S) contains 5
sources with SEDs that are well-fit as YSOs.  Five relatively faint
24\,{\um} sources that are likely to be YSOs are also scattered
throughout this region.  Sources in this southern region are more
distributed and are not spatially well-correlated with the dark cloud
filament.  This is in contrast to the YSOs in the northern region
where they follow the cloud boundaries reasonably well.  Mass
estimates range from about 1 to 7\,M{\sun}.

\section{Discussion}

The IR-dark cloud associated with IRAS\,18507$+$0121 has a filamentary
structure that spans more than $9'$ on the sky ($> 10$\,pc at
3.9\,kpc).  Splitting the cloud into three regions (south, central and
north) one finds that the southern region is associated with mostly
low to intermediate-mass YSO candidates (10 in total) that are not
spatially well-correlated with the dark cloud filament.  The northern
segment of the filament contains 23 low to intermediate-mass YSO
candidates that tend to trace the IR-dark cloud structure. The central
region of the filament is in the process of forming four or five mid-
to early-B stars with $M_\star > 8$\,M{\sun}.  Four more sources are
likely low-to-intermediate mass with $M_\star$ between 0.5 and
8\,M{\sun}.  Together with an additional 11 YSO candidates seen at
24\,{\um} this central region harbors a relatively compact cluster
with a total of 20 YSO candidates.

\subsection{GLIMPSE-detected source characteristics}

Bright emission in the 4.5\,$\mu$m Spitzer IRAC can be a strong
indicator of molecular outflow (e.g. Qiu et al. 2007, Smith et
al. 2006, Noriega-Crespo et al. 2004).  The most likely reason for
this is the presence of H$_2$ emission lines and/or emission from the
CO 1-0 P(8) bandhead.  All four IRAC bands contain H$_2$ emission
lines, but these features are narrow and hence can only significantly
enhance the brightness in the broadband images if the lines are
extremely bright relative to the continuum.  Because {\h} features are
distributed throughout the IRAC bands, strong {\h} lines will not
cause a greater enhancement of the flux in band [4.5] than in bands
[5.8] and [8.0] (see e.g., Smith et al. 2006).  In contrast, the CO 1-0
P(8) emission bandhead is a broad feature that occupies the entire
[4.5] filter and can become bright in the presence of shocked,
molecular gas.  There is not significant CO bandhead emission in the
remaining IRAC bands.  Thus, sources with {\it excess} emission in
band [4.5] relative to bands [5.8] and [8.0] appear likely to have CO
bandhead emission resulting from shocked, molecular gas.  This does
not preclude the possibility that significant {\h} emission may also
contribute to the brightness in all IRAC bands.

Three of the 29 YSOs listed in Table 3 that were detected in all four
IRAC bands have noticeable excess emission at 4.5{\um} and strong flux
at 24{\um}: sources 6 \& 12 in region C, and source 29 in region N.
Because of this, SED fits used the [4.5] fluxes as upper limits only
since the YSO SED models do not incorporate molecular line emission.
These sources are likely to have significant shocked, molecular gas
and are undergoing outflow \& accretion.  Source 31, although it was
not well-fit, should probably be considered in this group as well.  It
was not well-fit because it has a strong 4.5{\um} excess and it was
only detected in the four IRAC bands.  The SED (not shown) looks
similar to G34.4\,MM itself.  Further, its non-detection at 24{\um} is
almost certainly due to its proximity to G34.4\,MM.

The most prominent of the 4.5{\um} excess sources is source 12, the
mid-IR counterpart to G34.4 MM and outflow G34.4:A.  The association
of this source with the CO outflow favors the interpretation that the
4.5{\um} excess is due to shocked CO emission.

The potential drivers of outflows G34.4:C (source 11) and G34.4:E
(source 10) do not show similar 4.5{\um} excess fluxes.  A possible
reason for this may be that they are somewhat more evolved than source
12.  In very young sources, particularly those with $M_\star > 2$
\Msun, high accretion rates create a cool, distended photosphere that
does not emit strongly enough in the UV to excite the PAH emission
features present in the IRAC [3.6], [5.8] and [8.0] bands.  In more
evolved sources, however, PAH emission can hide the contrast between
[4.5] and the other 3 bands.  Alternatively, as discussed above, both
sources 10 and 11 suffer from confusion with the {\uchii} region which
emits strongly in all four bands and this could also mask moderate
enhancements due to CO bandhead emission.

As a final note, three sources show marginal excess at 4.5{\um}
relative to bands [5.8] and [8.0]: source 7 in region C and sources 20
\& 26 in region N.  These sources do not appear to have strong
emission at 24{\um}, so they may be a different type of source than
those with strong 4.5 \& 24{\um} excesses.

\subsection{Timescale for Massive Star Formation}

There are two cloud cores in the central region of the filament: one
more massive core in the south surrounded by a more diffuse molecular
cloud that contains the {\uchii} region and one that is more compact
containing source 12/G34.4\,MM.  The cores are separated by about
0.75\,pc.  Each cloud core is producing a small cluster of massive
flows that look similar and are of similar ages although the northern
outflows associated with G34.4\,MM have about one third the mass as
the southern flows.

Due to the shorter formation timescales ($\ltsim 10^5$ years), massive
(proto)stars begin to burn hydrogen even while they continue to
accrete material.  The generation of a {\uchii} region identifies
which sources have reached the main sequence and begun hydrogen
burning while the detection of a molecular outflow traced to the
source identifies whether it is still accreting (and hence still
powering an outflow).  However, the addition of material onto the
stellar surface implies that the source will move up the main
sequence.  In this sense it is still a {\it protostar} because it has
not reached its final mass.

The central source in the G34.4 {\uchii} region (source 11) has a
spectral type of B0.5 while the spectral type of source 12/G34.4\,MM
is probably closer to a B2 star (this work \& SNB04). High-velocity CO
emission can be traced to both sources suggesting that they are still
accreting even though they have reached the main-sequence.

The characteristic time scale of the outflow/infall phase from early B
stars is roughly a few $\times 10^5$ years (Richer et al. 2000) while
the evolutionary timescale for a B0.5 star to reach the main-sequence
is roughly $10^5$ years.  Since ionized gas is seen toward both cores
and the outflow characteristics are similar, it seems reasonable to
assume that they began forming massive stars at approximately the same
time: about $10^5$ years ago.

SNB04 detected a lower mass population of this cluster (e.g. a few
solar masses or less) at near-infrared wavelengths (see their Fig. 6).
Their detected cluster population was more distributed, less embedded
and less massive than those detected in this work. They found that
this sub-population is less than 3\,Myrs old but probably more than
0.3\,Myrs old.

There are numerous problems associated with these age estimates and
they could be in error easily by a factor of a few.  However, let us
take them at face value for now.  SNB04 discovered a distributed
population of low to intermediate mass stars that may have formed a
few Myrs ago.  The more massive stars detected in this work are more
tightly associated with the dense molecular gas (as expected) and
appear to be somewhat younger, more like $10^5$ years old.  This
suggests that the star formation in this cold, dark cloud may have
formed in two stages: first lower mass stars began to form, then, a
few Myrs later, the more massive, early B stars began to form.

Such a situation is predicted when clouds have magnetic fields that
are stronger than needed to provide support against gravitational
collapse and the core mass to magnetic flux ratio (M/$\Phi$) is
sub-critical.  Thus, low-mass stars would form in originally highly
magnetically sub-critical clouds, with ambipolar diffusion leading to
core formation, quasi-static contraction of the cores, and eventually,
massive star formation (see e.g. the review by Ward-Thompson et al
(2007) and references therein).  If this scenario is correct, then the
cold, dark filament associated with IRAS 18507 may provide an example
of a cloud with delayed massive star formation due to an initially
high magnetic field and sub-critical M/$\Phi$.

\subsection{Star Formation Efficiency}

The set of best-fit SED models for YSO candidates can be used to
construct a $\chi^2$-weighted probability distribution of the best-fit
model parameters.  We sum the mass probability distributions for each
source to construct mass functions for all well-fit YSOs (Fit=1) in
Tables 3 \& 4.  The total mass of the GLIMPSE-detected YSOs in this
IR-dark cloud is then $127\pm 27$\,\Msun.  

The total stellar mass of the central (sub)cluster containing
G34.4\,MM and the \uchii\ region is $48\pm 20$\,\Msun (or $\sim 38$\%
of the total detected stellar mass).  About 14\,{M\sun} is in source
12/G34.4\,MM, 10\,M{\sun} is in source 31 (the yellow source near
G34.4\,MM in Fig. 10), and the remaining 24\,{M\sun} of stellar mass
resides in the central star producing the {\uchii} region and its
surrounding cluster.  These rough stellar mass estimates are
incomplete for $M_\star < 3$\,\Msun.  They also do not include the
additional 20 YSO candidates that were detected at 24\,\um\ but in
fewer than three IRAC bands.  Thus, the stellar mass estimate should
be considered a lower limit.

Using the initial mass function (IMF) of Kroupa et al. (2001) (where
the IMF slope follows a Salpeter IMF for $M_\star >1$\,{M\sun} and it
becomes shallower for $0.1 < M_\star < 1$\,{M\sun}) one can estimate
the expected stellar mass for $M_\star < 3$\,\Msun.  Assuming $127\pm
27$\,{M\sun} of mass in stars more massive than 3\,{M\sun}, roughly
$413\pm 103$\,{M\sun} of mass is expected to be in the form of stars
less massive than 3\,{M\sun}.  Thus, the total stellar mass content of
this cloud is expected to be $\sim 540 \pm 130$\,{M\sun}.

This stellar mass estimate is a little more than half of the derived
mass of the ammonia hot core of 1000\,{\Msun} surrounding the {\uchii}
region (Maralles, Rodr{\'{\i}}gues \& Scalise 1994).  And it is about
7\% of the cloud mass of 7500\,{\Msun} derived from 1.2\,mm
continuum emission (Rathborne et al. 2005).  The total star formation
efficiency is then $M_\star / (M_{cloud} + M_\star) = 540/(8040) \sim
0.07$ or 7\%.  This should be considered an upper limit because
1.2\,mm continuum emission traces warm dust (typically 30-50\,K) which
is heated predominantly by massive stars - it does not effectively
trace the cold gas and dust (5-10\,K) which may represent the bulk of
the total cloud mass.

We can also obtain an estimate of the massive and intermediate mass
star formation efficiency in the warm gas traced by 1.2\,mm continuum
emission and in the densest part of the hot core.  Assuming a warm
cloud mass of 7500\,{\Msun} and a total mass of stars with $M_\star >
3$\,\Msun$= 127$ \,{M\sun}, the massive star formation efficiency is
roughly 2\%.

\subsection{Can the outflows destabilize the cloud cores? }

\noindent
{\bf The G34.4 cloud core surrounding the \uchii\ region:} \\
Maralles, Rodr{\'{\i}}gues \& Scalise (1994) estimated a mass of the
G34.4 cloud core surrounding the \uchii\ region to be $\sim
1000$\,M{\sun} based on NH$_3$ observations.  They only detect
emission near the UC\,H{\sc ii} region, not near the G34.4\,MM core.
The {\ceo} mass estimate of 693\,M{\sun} compares reasonably well with
that of NH$_3$ given the uncertainties inherent with both estimates
and the fact that our {\ceo} map is missing extended emission.

If we take the mass of the G34.4 core surrounding the \uchii\ region
to be $M_{cloud} = 1000$\,M{\sun} we find that the core has about nine
times more mass than in the high-velocity molecular outflows.  The
gravitational binding energy of the cloud, can be calculated from $G
M_{cloud}^2/c_1 r$, where $c_1$ is a constant which depends on the
mass distribution ($c_1 = 1$ for $\rho \propto r^{-2}$).  We find that
the gravitational binding energy is $2.4 \times 10^{47}$~ergs.
Roughly 11\% of the molecular cloud core is participating in the
outflow and the combined outflow energy is roughly one third the
gravitational binding energy of the cloud.  The G34.4 outflows near the
UC\,H{\sc ii} region are injecting a significant amount of mechanical
energy into the cloud core and may help prevent further collapse of
the cloud.  However, they are not likely to totally disrupt the cloud.
As long as cloud material remains, this leaves the door open for
possible future episodes of star formation.

\noindent
{\bf The compact G34.4\,MM cloud core:} \\ 
SNB04 estimate the mass of the G34.4\,MM core based on 2.6 millimeter
thermal dust emission to be between 150 and 650\,M{\sun} with
240\,M{\sun} being the value derived for an opacity index, $\beta$ of
1.5 and a temperature of 50\,K.  Our estimated mass based on {\ceo}
emission is significantly less than this ($\sim 75$\,M\sun) suggesting
that our {\ceo} maps are missing extended emission or the assumptions
used to estimate the mass associated with warm dust emission in the
G34.4\,MM core were not correct.  Thus, for example, if $\beta = 0.7$
and $T_d = 70$\,K then the mass of the millimeter core would be $\sim
75$\,M{\sun} which is consistent with the mass derived from {\ceo}.
Such a low opacity index and high temperature could be caused by the
presence of strong shocks through most of the core mass which would
prevent large grains from forming and maintain the bulk of the gas at
higher temperatures.  Indeed, this could be possible given the small
size of the core (0.2\,pc) and the fact that two outflows with a
combined mass of nearly 35\,M{\sun} are blasting their way out of the
core.

If we assume a mass of 75-240\,M{\sun} for the G34.4\,MM core, as
derived from the {\ceo} and 2.6\,mm continuum observations, then the
gravitational binding energy of the cloud is $5.6 \times 10^{45}$\,ergs
to $5.7 \times 10^{46}$\,ergs.  The kinetic energy of the outflowing
gas is between 1.2 to nearly nine times the gravitational binding energy
of the cloud core.  Thus, the G34.4\,MM flows appear to have more
energy relative to the core and are more likely to disrupt the compact
core.  Therefore we may be seeing the first and only episode of
massive star formation in this compact region.

\section{Summary}

We have observed the central region of the IR-dark cloud associated
with IRAS\,18507$+$0121 in CO J=1--0, {\tcof} and {\ceof} with the
Owens Valley Radio Observatory at $\sim 4''$ resolution.  Five massive
outflows from two cloud cores were discovered.  A total of 146\,M\sun\
of material is participating in the combined outflows and injecting
$\sim 1.3 \times 10^{45}$\,ergs of kinetic energy.  We also used
archived Spitzer data from the GLIMPSE survey to gain an understanding
of the stellar content of the cluster and the entire IR-dark filament.
Modeling of the SEDs of GLIMPSE sources identified 31 young stellar
objects in the filament with a combined stellar mass of $\sim 127 \pm
27$\,\Msun.  An additional 22 sources were identified as probable
cluster members based on the presence of strong 24\,\um\ emission.

The total star formation efficiency in the G34.4 cloud filament is
estimated to be about 7\% while the massive and intermediate mass
star formation efficiency in the entire cloud filament is estimated to
be roughly 2\%.

A qualitative evaluation of the outflow characteristics and presence
of ionized gas in the central region of the IR-dark cloud filament
suggests that the early B stars currently undergoing outflow and
accretion began forming stars at approximately the same time: about
$10^5$ years ago.  A population of lower mass stars were detected by
SNB04 that appear to have an age of about a few Myrs.  This dual
population suggests that stars in the IRAS 18507 cloud may have formed
in two stages: first lower mass stars formed and then, a few Myrs
later, the more massive, early B stars began to form.  Such a
situation can occur in magnetically sub-critical clouds in which slow,
ambipolar diffusion leads to the delayed onset of core contraction and
eventual massive star formation.

A comparison of the gravitational binding energy with the outflow
kinetic energy in the two cloud cores studied in molecular lines
suggests that the compact core containing G34.4\,MM is being destroyed
by its molecular outflows.  The outflows have as much or more kinetic
energy than the gravitational binding energy of the cloud core.  Thus,
we may be seeing the first and only episode of massive star formation
in this compact region.

In contrast, the 690\,\Msun\ core surrounding the G34.4 \uchii\ region
has twice the gravitational binding energy as the combined outflow
kinetic energy.  Thus, although the outflows are injecting a
significant amount of mechanical energy into the cloud core and may
help prevent further collapse of the cloud, they are not likely to
totally disrupt the cloud.  As long as cloud material remains, this
leaves the door open for possible future episodes of star formation.

\noindent {\bf Acknowledgments}\\ 
DSS would like to thank Allison Sills for useful discussions on
properties of young clusters.  Research at the Owens Valley Radio
Observatory is supported by the National Science Foundation through
NSF grant number AST 96-13717.  Star formation research at Owens
Valley is also supported by NASA's Origins of Solar Systems program,
Grant NAGW-4030, and by the Norris Planetary Origins Project.  LB
acknowledges support from the Chilean Center for Astrophysics FONDAP
grant 15010003. Support for GLIMPSE, part of the Spitzer Space
Telescope Legacy Science Program, was provided by NASA through
contracts 1224653 (University of Wisconsin at Madison) and 1224988
(Space Science Institute).  BW gratefully acknowledges support from
the NASA Astrophysics Theory Program (NNG05GH35G) and the Spitzer
Space Telescope Theoretical Research Program (Subcontract 1290701).

\clearpage
\begin{table}
\caption[]{G34.4 and G34\,MM Outflow Parameters}
\label{tab:par}
\smallskip
\begin{tabular}{|lll|}
\hline
Source:             & G34.4 combined flows       
                    & G34\,MM combined flows \\
\hline
\hline
CO radius of outflow  
                    &0.98 pc    &$0.85$ pc         \\
Assumed inclination angle  
                    &45\deg\   &45\deg\              \\
Outflow Mass$^{\dagger}$ :       
                    &          &              \\
~~~~Red-shifted: $^{12}$CO                        
                    &22.6 M$_\odot$             
                    &10.6 M$_\odot$         \\
~~~~~~~~~~~~~~~~~~~~~{\tco}                        
                    &59.6 M$_\odot$             
                    &12.7 M$_\odot$            \\
~~~Blue-shifted: $^{12}$CO
                    &14.2 M$_\odot$ 
                    &1.6 M$_\odot$ \\
~~~~~~~~~~~~~~~~~~~~~{\tco} 
                    &\underline{14.8 M$_\odot$} 
                    &\underline{9.9 M$_\odot$} \\
~~
                    & 111.2 M$_\odot$           
                    & ~34.8 M$_\odot$         \\
Momentum 
                    &$8.3 \times 10^2$ M$_\odot$ km s$^{-1}$  
                    &$3.5 \times 10^2$ M$_\odot$ km s$^{-1}$  \\
Kinetic Energy     
                    &$7.8 \times 10^{46}$ ergs    
                    &$4.9 \times 10^{46}$ ergs  \\
Dynamical time scale of CO flow
                    &$\sim 4 \times 10^4$ yr      
                    &$\sim 3.8 \times 10^4$ yr   \\
$\dot M_f$ 
                    &$1.3 \times 10^{-3}$ M$_\odot$ yr$^{-1}$ 
                    &$0.8 \times 10^{-3}$ M$_\odot$ yr$^{-1}$   \\
Momentum Supply Rate (Force) 
                    &$12.2 \times 10^{-3}$ M$_\odot$ km s$^{-1}$ yr$^{-1}$ 
                    &$~8.0 \times 10^{-3}$ M$_\odot$ km s$^{-1}$ yr$^{-1}$ \\
Mechanical Luminosity 
                    &11.2  L$_\odot$             
                    &~9.1  L$_\odot$ \\
\hline
\end{tabular}

\vspace{3mm}
~~{\small ${\dagger}$ $^{12}$CO outflow emission (corrected for
  optical depth from [CO]/[{\tco}]) \\

\vspace{-6mm}
~~~~~~~~~~ measured between velocities 32.9 to 51.7\kms\ and 63.6 to
            77.8\kms. \\ 

\vspace{-5mm}
~~~~{\tco} outflow emission (assumed to be optically thin) \\

\vspace{-6mm}
~~~~~~~~~~ measured between velocities 51.7 to 53.6\kms\ and 60.4 to
            63.6\kms.} \\  

\end{table}

\begin{table}
\caption[]{G34.4 and G34\,MM Core Parameters measured from {\ceo}}
\label{tab:par}
\smallskip
\begin{tabular}{|lll|}
\hline
Source:           & G34.4       
                    & G34\,MM \\
\hline
\hline
{\ceo}
Core Mass:        &692.7 M$_\odot$        
                    &75.2 M$_\odot$              \\
$v_{LSR}$         &56.0{\kms}
                    &57.4{\kms}         \\
Momentum          &$2.0 \times 10^3$ M$_\odot$ km s$^{-1}$  
                    &$2.5 \times 10^2$ M$_\odot$ km s$^{-1}$  \\
Kinetic Energy    &$8.0 \times 10^{46}$ ergs    
                    &$1.2 \times 10^{46}$ ergs  \\
\hline
\end{tabular}
\end{table}

\clearpage
\begin{deluxetable}{rccrrrrrrrrrrcc}
  \tabletypesize{\scriptsize}
  \rotate
  \tablecaption{YSO Cluster Members}
  \tablewidth{0pt}
  \tablehead{
   \colhead{} & \multicolumn{2}{c}{Coordinates} &
   \multicolumn{10}{c}{Measured Mid-IR Fluxes and
      Uncertainties (mJy)} & \multicolumn{2}{c}{} \\
    \colhead{Src.\tablenotemark{a}} & \colhead{$\alpha$(J2000)} &
    \colhead{$\delta$(J2000)} & \colhead{$F[3.6]$} & \colhead{$\delta
      F[3.6]$} & \colhead{$F[4.5]$} & \colhead{$\delta F[4.5]$} &
    \colhead{$F[5.8]$} & \colhead{$\delta F[5.8]$} &
    \colhead{$F[8.0]$} & \colhead{$\delta F[8.0]$} & \colhead{$F[24]$}
    & \colhead{$\delta F[24]$\tablenotemark{b}} & \colhead{Fit\tablenotemark{c}} & \colhead{Reg.\tablenotemark{d}} 
}
\startdata
  1 & \ra{18}{53}{24.999} &  \dec{1}{23}{01.4} &    3.43 &    0.34 &    2.11 &    0.23 &    1.86 &    0.37 &     --- &     --- &     1.61  &     0.16  & 1 & S \\
  2 & 18 53 18.649 &  1 23 21.8 &    1.78 &    0.18 &    2.01 &    0.20 &    2.85 &    0.28 &    2.68 &    0.40 &    10.54  &     1.68  & 1 & S \\
  3 & 18 53 16.833 &  1 23 26.6 &    5.20 &    0.52 &    6.10 &    0.61 &    5.07 &    0.51 &    4.33 &    0.65 &     6.73  &     0.67  & 1 & S \\
  4 & 18 53 19.694 &  1 23 36.2 &    6.78 &    0.68 &   16.79 &    1.68 &   28.13 &    2.81 &   20.32 &    3.05 &    60.00  & U.L. &  1 & S \\
  5 & 18 53 17.773 &  1 23 49.4 &    8.66 &    0.87 &   12.26 &    1.23 &   13.92 &    1.39 &   15.76 &    2.36 &    45.29  &     4.53  & 1 & S \\
  6 & 18 53 18.982 &  1 24 11.5 &    0.84 &    0.15 &    7.66 &    0.93 &    5.00 &    0.86 &    2.79 &    0.64 &   307.85  &    30.78  & 1 & C \\
  7 & 18 53 18.825 &  1 24 19.6 &   12.22 &    1.22 &   26.14 &    2.61 &   26.69 &    2.67 &   23.15 &    3.47 &   150.00  & U.L. &  1 & C \\
  8 & 18 53 19.404 &  1 24 24.2 &    0.25 &    0.07 &    1.46 &    0.16 &    3.33 &    0.36 &    4.18 &    0.63 &   150.00  & U.L. &  1 & C \\
  9 & 18 53 21.127 &  1 24 27.5 &    3.79 &    0.38 &    3.56 &    0.36 &    3.07 &    0.52 &    2.00 &    0.30 &    19.92  &     2.54  & 1 & C \\
 10\tablenotemark{e} & 18 53 18.727 &  1 24 43.0 &    5.13 &    0.92 &   19.57 &    1.96 &   40.99 &    6.81 &   59.72 &    8.96 & 18,024.00  & U.L. &  1 & C \\
 11\tablenotemark{e} & 18 53 18.630 &  1 24 48.3 &   28.26 &    7.77 &   79.15 &   14.59 &  168.90 &   16.89 &  277.30 &   45.02 & 18,024.00  & U.L. &  1 & C \\
 12\tablenotemark{f} & 18 53 18.058 &  1 25 25.3 &    2.24 &    0.22 &   16.20 &    8.00 &   13.85 &    1.38 &   11.34 &    1.13 &  8,682.80  &  868.28  & 1 & C \\
 13 & 18 53 16.533 &  1 26 28.6 &    0.37 &    0.06 &    1.03 &    0.10 &    1.40 &    0.22 &    1.59 &    0.24 &     7.11  &     0.71  & 1 & N \\
 14 & 18 53 17.321 &  1 26 33.0 &    1.88 &    0.19 &    2.65 &    0.27 &    2.83 &    0.40 &    1.77 &    0.27 &    15.01  &     1.50  & 1 & N \\
 15 & 18 53 17.232 &  1 26 41.1 &    5.01 &    0.50 &    6.90 &    0.69 &    7.18 &    0.72 &    6.23 &    0.93 &    11.21  &     1.12  & 1 & N \\
 16 & 18 53 29.815 &  1 26 52.7 &    3.30 &    0.33 &    4.12 &    0.41 &    4.68 &    0.53 &    5.80 &    0.87 &     8.09  &     0.81  & 1 &  \\
 17 & 18 53 21.571 &  1 26 58.9 &    0.38 &    0.06 &    1.58 &    0.23 &    1.58 &    0.35 &     --- &     --- &    13.46  &     1.61  & 1 & N \\
 18 & 18 53 11.979 &  1 27 09.5 &   13.41 &    1.34 &   10.45 &    1.04 &    8.31 &    0.83 &    7.87 &    1.61 &   119.71  &    37.16  & 1 & N \\
 19 & 18 53 17.787 &  1 27 12.6 &    0.81 &    0.09 &    2.11 &    0.21 &    3.29 &    0.33 &    3.94 &    0.59 &    13.88  &     1.39  & 1 & N \\
 20 & 18 53 18.181 &  1 27 28.5 &    1.40 &    0.19 &    3.23 &    0.32 &    3.30 &    0.41 &    3.40 &    0.51 &    13.21  &     1.32  & 1 & N \\
 21 & 18 53 15.807 &  1 27 37.4 &   40.59 &    4.06 &   41.08 &    4.11 &   46.97 &    4.70 &   36.85 &    5.53 &    20.79  &     2.08  & 1 & N \\
 22 & 18 53 18.604 &  1 27 38.8 &    0.49 &    0.09 &    0.71 &    0.11 &     --- &     --- &    2.66 &    0.40 &    10.97  &     1.40  & 1 & N \\
 23 & 18 53 23.391 &  1 27 57.0 &    2.47 &    0.25 &    2.52 &    0.25 &    2.70 &    0.34 &    2.35 &    0.35 &     4.07  &     0.41  & 1 & N \\
 24 & 18 53 23.784 &  1 28 01.1 &    0.49 &    0.08 &    0.55 &    0.06 &    1.02 &    0.32 &    0.96 &    0.14 &     4.94  &     0.49  & 1 & N \\
 25 & 18 53 18.848 &  1 28 02.7 &    0.65 &    0.09 &    0.58 &    0.09 &     --- &     --- &    2.59 &    0.39 &      ---  &      ---  & 1 & N \\
 26 & 18 53 19.404 &  1 28 16.1 &    0.83 &    0.09 &    2.12 &    0.21 &    2.60 &    0.42 &    2.93 &    0.44 &     4.16  &     0.42  & 1 & N \\
 27 & 18 53 20.005 &  1 28 19.6 &    0.55 &    0.08 &    1.29 &    0.23 &    2.14 &    0.68 &    1.47 &    0.22 &    29.57  &     2.96  & 1 & N \\
 28 & 18 53 19.049 &  1 28 21.5 &    2.53 &    0.29 &    2.64 &    0.26 &    2.23 &    0.24 &    1.68 &    0.25 &     5.42  &     0.54  & 1 & N \\
 29 & 18 53 20.674 &  1 28 24.1 &    9.65 &    0.96 &   32.79 &    3.94 &   26.83 &    2.68 &    8.03 &    1.20 &   278.27  &    27.83  & 1 & N \\
 30 & 18 53 20.078 &  1 28 43.1 &    0.27 &    0.05 &    1.02 &    0.11 &    1.83 &    0.28 &    2.49 &    0.37 &    15.80  &     1.58  & 1 & N \\
 31 & 18 53 18.586 &  1 25 38.9 &    0.33 &    0.08 &    3.43 &    0.39 &    2.50 &    0.32 &    1.79 &    0.27 &      ---  &      ---  & 2 & C \\
 32 & 18 53 19.144 &  1 24 35.4 &    0.68 &    0.09 &    1.46 &    0.18 &     --- &     --- &     --- &     --- &      ---  &      ---  & 3 & C \\
 33 & 18 53 13.870 &  1 26 03.3 &    0.98 &    0.13 &    0.71 &    0.11 &     --- &     --- &     --- &     --- &     3.31  &     0.59  & 4 & N \\
 34 & 18 53 13.933 &  1 24 46.7 &     --- &     --- &     --- &     --- &     --- &     --- &     --- &     --- &     5.79  &     0.40  & 4 &  \\
 35 & 18 53 14.135 &  1 25 03.8 &    0.35 &    0.06 &    0.34 &    0.06 &     --- &     --- &     --- &     --- &     5.52  &     1.00  & 4 &  \\
 36 & 18 53 14.702 &  1 25 25.0 &     --- &     --- &     --- &     --- &     --- &     --- &     --- &     --- &     8.00  &     0.99  & 4 &  \\
 37 & 18 53 14.884 &  1 25 49.0 &     --- &     --- &     --- &     --- &     --- &     --- &     --- &     --- &     4.56  &     0.50  & 4 &  \\
 38 & 18 53 15.298 &  1 25 48.1 &    1.27 &    0.15 &    0.86 &    0.09 &     --- &     --- &     --- &     --- &    11.14  &     1.11  & 4 &  \\
 39 & 18 53 15.562 &  1 23 25.2 &     --- &     --- &     --- &     --- &     --- &     --- &     --- &     --- &     2.75  &     0.53  & 4 & S \\
 40 & 18 53 16.029 &  1 24 25.4 &     --- &     --- &     --- &     --- &     --- &     --- &     --- &     --- &    21.63  &     1.27  & 4 & C \\
 41 & 18 53 16.811 &  1 24 11.9 &     --- &     --- &     --- &     --- &     --- &     --- &     --- &     --- &    10.22  &     0.47  & 4 & C \\
 42 & 18 53 16.908 &  1 26 32.0 &     --- &     --- &     --- &     --- &     --- &     --- &     --- &     --- &     8.87  &     0.92  & 4 & N \\
 43 & 18 53 18.026 &  1 26 25.9 &     --- &     --- &     --- &     --- &     --- &     --- &     --- &     --- &    42.84  &     2.55  & 4 & N \\
 44 & 18 53 18.065 &  1 23 24.8 &     --- &     --- &     --- &     --- &     --- &     --- &     --- &     --- &     7.76  &     0.97  & 4 & S \\
 45 & 18 53 18.085 &  1 26 12.0 &    0.49 &    0.07 &    0.45 &    0.08 &     --- &     --- &     --- &     --- &     8.27  &     3.97  & 4 & N \\
 46 & 18 53 18.409 &  1 23 38.7 &     --- &     --- &     --- &     --- &     --- &     --- &     --- &     --- &    10.10  &     0.83  & 4 & S \\
 47 & 18 53 18.516 &  1 23 56.8 &     --- &     --- &     --- &     --- &     --- &     --- &     --- &     --- &    17.53  &     4.67  & 4 & C \\
 48 & 18 53 18.589 &  1 26 49.5 &     --- &     --- &     --- &     --- &     --- &     --- &     --- &     --- &     4.47  &     0.50  & 4 & N \\
 49 & 18 53 18.600 &  1 23 46.2 &     --- &     --- &     --- &     --- &     --- &     --- &     --- &     --- &    20.59  &     2.42  & 4 & S \\
 50 & 18 53 19.904 &  1 28 08.6 &     --- &     --- &    0.68 &    0.08 &    1.30 &    0.36 &     --- &     --- &    12.96  &     1.30  & 4 & N \\
 51 & 18 53 21.233 &  1 25 12.2 &     --- &     --- &     --- &     --- &     --- &     --- &     --- &     --- &    14.51  &     0.83  & 4 & C \\
 52 & 18 53 22.760 &  1 22 30.3 &     --- &     --- &     --- &     --- &     --- &     --- &     --- &     --- &     6.44  &     0.95  & 4 & S \\
 53 & 18 53 22.826 &  1 25 10.0 &    0.80 &    0.10 &    0.67 &    0.10 &     --- &     --- &     --- &     --- &    35.49  & U.L. &  4 &  \\
 54 & 18 53 22.905 &  1 25 07.0 &    1.58 &    0.16 &    1.53 &    0.18 &     --- &     --- &     --- &     --- &    35.49  & U.L. &  4 &  \\
\enddata
\tablenotetext{a}{Sources are grouped in order of Fit flag, and
  then listed in order of increasing declination.}
\tablenotetext{b}{Where 2 or more GLIMPSE sources are confused with a
  single MIPS source, we treat the 24 \um\ flux as an upper limit (U.L)
  only.}
\tablenotetext{c}{The Fit flags are designated as follows: \\
  ~~~~~~1 =
  Source well-fit as a YSO, meeting the $\chi^2/N_{\rm data} \leq 4.8$
  criterion discussed in the text.\\
  ~~~~~~2 = Fit failed to meet the $\chi^2$ criterion, but source is
  considered to 
  be a probable YSO based upon visual inspection of its mid-IR SED. \\
  ~~~~~~3 = Not able to fit because source is detected only in the
  GLIMPSE [3.6] \& [4.5] bands (includes the possible source of
  Outflow G34.4:D). \\ 
  ~~~~~~4 = MIPS 24-$\mu$m-detected source not
  fittable because it is detected in fewer than 3 of the IRAC+2MASS
  bands. The majority of these sources are probably YSOs due to their
  location near the IR dark cloud, but a few could be field stars on
  the asymptotic giant branch.}
\tablenotetext{d}{Region of Figure 11 in which the
  source is located. Region C refers to the central part of the IR
  dark cloud containing
  the OVRO observations. Region N represents the dark cloud north of C, while
  Region S represents the dark cloud south of C. Unlabeled sources
  lie outside of these 3 regions.}
\tablenotetext{e}{These sources overlie the UC H{\scshape ii} region,
  which produces bright diffuse emission in all 4 {\it MSX} bands as well as
  at MIPS 24 $\mu$m and 70 $\mu$m. 
  Although these YSOs may be bright enough to be detected
  by {\it MSX} and MIPS, they are confused
  with the UC H{\scshape ii}.}
\tablenotetext{f}{The IR counterpart to G34.4\,MM. It is
  detected also by {\it MSX} at 21.3 $\mu$m and by MIPS at 70 $\mu$m, 
  with fluxes of 4,900$\pm 258$ and 322,000$\pm 128,000$ mJy,
  respectively.}
\end{deluxetable}

\clearpage
\begin{deluxetable}{cccccrrrrr}
  \tabletypesize{\footnotesize}
  \rotate
  \tablecaption{YSO Model Parameters\tablenotemark{a} from SED Fits to Cluster Members}
  \tablewidth{0pt}
  \tablehead{
    \colhead{Src.} & \colhead{Fit}  & 
    \colhead{$\chi^2_{min}$} &
    \colhead{$\chi^2_{max}$\tablenotemark{b}} & \colhead{\# Fits} & \colhead{$A_{V}$} &
    \colhead{$M_{\star}$ (\Msun)} & \colhead{$L_{\star}$
      (L$_{\odot}$)} & \colhead{$\dot{M}_{\rm env}$ (\Msun\
      yr$^{-1}$)\tablenotemark{c}} & \colhead{$i$}
}
\startdata
  1 & 1 &  7.63 & 13.57 &  112 &  8.5 --  15.1 & 2.63E+00 -- 4.34E+00 & 3.44E+01 -- 2.83E+02 &       0 & 18.2 -- 87.1 \\
  2 & 1 &  1.82 &  7.81 &  322 &  1.5 --  24.7 & 5.81E-01 -- 3.95E+00 & 6.22E+00 -- 4.11E+01 &       0 -- 3.44E-05 & 18.2 -- 81.4 \\
  3 & 1 &  8.24 & 14.21 &   96 & 11.4 --  22.8 & 2.65E+00 -- 4.24E+00 & 4.82E+01 -- 1.18E+02 &       0 -- 3.06E-07 & 41.4 -- 81.4 \\
  4 & 1 &  8.18 & 12.08 &   29 & 22.5 --  58.3 & 1.03E+00 -- 6.77E+00 & 1.10E+02 -- 1.54E+03 &       0 & 18.2 -- 81.4 \\
  5 & 1 &  0.86 &  7.85 &  185 & 11.4 --  26.5 & 3.32E+00 -- 5.03E+00 & 1.03E+02 -- 3.86E+02 &       0 -- 8.44E-06 & 18.2 -- 81.4 \\
  6 & 1 &  5.16 & 11.15 &   53 &  0.0 --   6.9 & 2.41E+00 -- 6.59E+00 & 8.22E+01 -- 3.40E+02 & 5.36E-05 -- 1.10E-03 & 31.8 -- 69.5 \\
  7 & 1 &  1.43 &  5.40 &  106 & 16.7 --  38.1 & 3.71E+00 -- 6.10E+00 & 1.82E+02 -- 9.68E+02 &       0 -- 1.60E-06 & 41.4 -- 81.4 \\
  8 & 1 &  2.83 &  6.82 &  323 &  0.0 --  59.9 & 1.94E-01 -- 5.18E+00 & 4.75E+00 -- 2.27E+02 &       0 -- 4.22E-04 & 18.2 -- 75.5 \\
  9 & 1 &  6.46 & 12.42 &   36 &  0.0 --  13.2 & 9.22E-01 -- 3.34E+00 & 9.97E+00 -- 3.53E+01 & 1.18E-06 -- 9.78E-05 & 49.5 -- 81.4 \\
 10 & 1 &  0.69 &  4.69 &  846 &  0.0 --  60.0 & 5.00E-01 -- 9.97E+00 & 4.47E+01 -- 4.46E+03 &       0 -- 1.04E-03 & 18.2 -- 81.4 \\
 11 & 1 &  0.17 &  5.17 &  879 &  5.2 --  50.2 & 2.13E+00 -- 9.92E+00 & 2.12E+02 -- 5.16E+03 &       0 -- 2.18E-03 & 18.2 -- 81.4 \\
 12\tablenotemark{d} & 1 &  6.62 & 36.13 &   10 & 38.8 --  41.5 & 1.36E+01 -- 1.39E+01 & 6.92E+03 -- 9.39E+03 & 5.91E-04 -- 9.96E-04 & 56.6 \\
 13 & 1 &  3.11 &  8.11 &  200 &  7.5 --  60.0 & 9.65E-01 -- 4.25E+00 & 1.12E+01 -- 1.49E+02 &       0 -- 1.01E-04 & 18.2 -- 87.1 \\
 14 & 1 &  1.57 &  6.36 &   31 &  0.0 --  18.7 & 4.33E-01 -- 2.70E+00 & 5.32E+00 -- 3.16E+01 & 7.22E-06 -- 1.10E-04 & 41.4 -- 81.4 \\
 15 & 1 &  0.78 &  5.77 &  186 &  0.0 --  25.5 & 1.46E+00 -- 4.40E+00 & 1.95E+01 -- 1.49E+02 &       0 -- 1.73E-05 & 18.2 -- 81.4 \\
 16 & 1 &  0.03 &  5.02 &  697 &  0.0 --  30.4 & 2.17E+00 -- 4.24E+00 & 1.67E+01 -- 1.97E+02 &       0 -- 6.68E-07 & 18.2 -- 81.4 \\
 17 & 1 &  7.49 & 11.49 &  470 & 15.8 --  60.0 & 2.04E+00 -- 5.44E+00 & 2.89E+01 -- 6.58E+02 &       0 -- 1.09E-07 & 18.2 -- 87.1 \\
 18 & 1 &  6.78 & 14.64 &   19 &  0.0 --   9.4 & 3.60E+00 -- 5.64E+00 & 1.10E+02 -- 1.66E+02 & 2.74E-05 -- 9.41E-04 & 18.2 -- 69.5 \\
 19 & 1 &  1.19 &  6.18 &  209 & 19.4 --  60.0 & 2.65E+00 -- 4.46E+00 & 2.89E+01 -- 3.14E+02 &       0 -- 2.82E-08 & 18.2 -- 81.4 \\
 20 & 1 &  4.89 & 11.30 &   14 &  0.0 --   7.1 & 6.60E-01 -- 1.54E+00 & 4.82E+00 -- 1.44E+01 &       0 -- 8.48E-06 & 18.2 -- 81.4 \\
 21 & 1 & 18.30 & 26.23 &   14 & 17.5 --  19.1 & 6.14E+00 -- 6.63E+00 & 9.96E+02 -- 1.34E+03 &       0 & 18.2 -- 87.1 \\
 22 & 1 &  0.01 &  4.01 &  425 &  1.5 --  60.0 & 4.18E-01 -- 4.71E+00 & 5.00E+00 -- 3.18E+02 &       0 -- 1.09E-04 & 18.2 -- 81.4 \\
 23 & 1 &  1.72 &  8.72 &  165 &  0.0 --  21.0 & 1.00E+00 -- 3.83E+00 & 1.04E+01 -- 5.35E+01 &       0 -- 1.88E-05 & 18.2 -- 81.4 \\
 24 & 1 &  0.62 &  5.62 & 1070 &  0.0 --  45.6 & 1.80E-01 -- 3.73E+00 & 1.44E+00 -- 5.09E+01 &       0 -- 2.99E-05 & 18.2 -- 81.4 \\
 25 & 1 &  0.24 &  4.23 &   22 &  6.7 --  10.5 & 1.89E+00 -- 2.77E+00 & 1.25E+01 -- 7.58E+01 &       0 & 18.2 -- 87.1 \\
 26 & 1 &  0.02 &  3.98 &  221 & 18.6 --  54.0 & 2.05E+00 -- 3.99E+00 & 1.67E+01 -- 2.05E+02 &       0 & 18.2 -- 87.1 \\
 27 & 1 &  0.39 &  5.38 &  293 &  0.0 --  52.3 & 4.14E-01 -- 5.03E+00 & 5.29E+00 -- 1.12E+02 & 2.95E-06 -- 3.31E-04 & 31.8 -- 81.4 \\
 28 & 1 &  3.85 & 11.84 &   98 &  3.6 --  14.5 & 7.91E-01 -- 3.64E+00 & 6.74E+00 -- 3.20E+01 &       0 -- 1.86E-05 & 41.4 -- 81.4 \\
 29 & 1 & 17.39 & 21.98 &    7 &  3.9 --  17.8 & 4.84E+00 -- 6.80E+00 & 1.69E+02 -- 3.42E+02 & 6.03E-05 -- 2.10E-04 & 41.4 -- 75.5 \\
 30 & 1 &  3.54 &  8.54 &  186 &  1.4 --  60.0 & 1.49E-01 -- 4.71E+00 & 1.99E+00 -- 2.92E+02 &       0 -- 1.71E-04 & 18.2 -- 81.4 \\
 31 & 2 & 39.45 & 51.32 &   19 &  7.4 --  40.5 & 7.51E+00 -- 1.34E+01 & 1.02E+03 -- 3.84E+03 & 2.67E-04 -- 2.21E-03 & 31.8 -- 87.1 \\
\enddata
\tablenotetext{a}{The range of values reported here for each YSO
  parameter has a 95\% probability, based on the probability distribution of best
  fits, of containing the actual value of that parameter for each source.}
\tablenotetext{b}{Fits having $\chi^2 \leq \chi^2_{\rm max}$ are used in
  computing the probability distribution of YSO parameters. For each
  source, $\chi^2_{\rm max}-\chi^2_{\rm min}=\Delta N_{\rm data}$,
  where $N_{\rm data}$ is
  the number of flux data points used in the fit and $\Delta = 1$ for
  Fit=1, $\Delta=3$ for Fit=2.}
\tablenotetext{c}{For models with $\dot{M}_{\rm env}<10^{-9}$ \Msun\
  yr$^{-1}$, the envelope accretion rate is considered to be zero.}
\tablenotetext{d}{For this source, the IR counterpart to G34MM, we
  report the probability distribution of parameters from the 10 best fits.}
\end{deluxetable}

\clearpage
\begin{center}  {\bf Figure Captions}  \end{center}

\noindent
{\bf Figure~1.~~} Four-color GLIMPSE image showing 3.6\,$\mu$m,
4.6\,$\mu$m, 5.8\,$\mu$m, and 8.0\,$\mu$m as blue, green, orange, and
red colors, respectively.  The image is displayed in galactic
coordinates and is roughly $16.6' \times 21.5'$ in size. The white box
at upper left gives the field of view shown in Figure 2.  The G34.4
UC\,H{\sc ii} region is in the center of the box within a dense cloud
filament seen as a dark lane. The mid-IR counterpart to the G34.4\,MM
core is seen as a faint green smudge immediately to the upper left of
the \uchii\ region.  The larger nebula in the lower right of the image
is associated with the cometary UC H{\sc ii} region G34.26. The bright
green emission emanating from this region traces a very energetic
outflow.  This GLIMPSE image includes the same data as reported in
Rathborne et al. (2005) but it covers a wider field-of-view to show
the physical relationship between the IR-dark cloud in which the G34.4
UC\,H{\sc ii} is located and the H{\sc ii} region G34.26.

\noindent
{\bf Figure~2.~~} Four images, all showing the same field of view.
Clockwise from the upper-left panel are: (a) 2MASS K-band, (b) GLIMPSE
4.5 \um, (c) GLIMPSE 8.0 \um, and (d) a 3-color image combining the
three other images as blue, green, and red, respectively.  Green plus
signs mark the positions of the UC\,H{\sc ii} and G34.4\,MM, while
green circles show the locations predicted for the driving sources of
outflows G34.4:D and G34.4:E. The white box gives the approximate
field of view of Figure 3.  The locations of the millimeter cores
detected by Rathborne et al. (2005) in J2000 coordinates are: MM1:
\ra{18}{53}{18.0}, \dec{01}{25}{24}; MM3: \ra{18}{53}{20.4},
\dec{01}{28}{23}; MM4: \ra{18}{53}{19.0}, \dec{01}{24}{08}; 

\noindent
{\bf Figure~3.~~} {\cof} red and blue-shifted emission.  Top left
panel shows red and blue-shifted emission from 62.3 to 79.2{\kms} and
37.5 to 53.2{\kms}, respectively.  The RMS is 470{\mjybkms} in the
red-shifted image and 370{\mjybkms} in the blue-shifted image.  Bottom
panels show low-velocity CO emission (left panel, 62.3 to 67.5{\kms}
and 49.3 to 53.2{\kms}, RMS = 350{\mjybkms}) and higher velocity CO
emission (right panel, 67.5 to 79.2{\kms} and 37.5 to 49.3{\kms}, RMS
= 350{\mjybkms}).  In all contour maps, contours are plotted at $3, 4,
5, 7, 10, 15, 20\, \sigma$ and continue with a spacing of $5 \sigma$.
The top right panel shows the NIR K band image of Shepherd et
al. (2004) with the locations of the proposed outflows overlaid.
%
%
In all panels: the solid circle represents the location of G34.4\,MM,
the star shows the location of the UC\,H{\sc ii} region, the outflow
(A) from G34.4\,MM is outlined with a series of arcs which delineate
the approximate boundaries of the flow.  Four other proposed outflows
are shown as arrows. A solid triangle indicates the possible location
of the driving source of an outflow.  The outflow from the UC\,H{\sc
ii} region is one-sided since only the blue-shifted outflow is clearly
detected.  The synthesized beam of $3.83'' \times 3.45''$ at
P.A. --59.7{\deg} is shown in the bottom right of the upper left
panel.

\noindent
{\bf Figure~4.~~} {\ceo} (top, left panel) and {\tco} (top, center
panel) integrated emission (moment 0) from 62.1\kms\ to 54.6\kms\ and
64.5\kms\ to 49.5\kms, respectively ($v_{LSR} = 57$\kms).  The RMS in
the {\ceo} image is 200\mjybkms; contours are plotted at $3, 5, 7, 9,
11, 13, 15, \& 17 \sigma$ while grey scale is shown from 200\mjybkms\
to a peak of 3.1\jybkms.  The RMS in the {\tco} image is 280\mjybkms;
contours are plotted at $3, 5, 7, 9, 12, 15, 20, 25,$ \& $30 \sigma$
while grey scale is shown from 280\mjybkms\ to a peak of 9.0\jybkms.
Plus symbols represent the locations of the UC\,H{\sc ii} region and
G34.4\,MM.  Green triangles represent the locations of spectra taken
for optical depth estimates in the outflows.  A scale size of 0.55 is
shown in the left panel, synthesized beams are shown in the bottom
right of each panel.
The top, right panel shows {\tco} red and blue-shifted emission from
63.1 to 61.8 {\kms} and 52.3 to 49.5{\kms}, respectively.
The RMS in the image is 100{\mjybkms}.  Contours are plotted at $3,
6, 9, 12, 15, 18$ and $21\, \sigma$.  Symbols are the same as in
Figure 3.  The bottom two panels show moment 1 maps of {\ceo} \&
{\tco} in which the brightness of the color is proportional to the
intensity in the moment 0 image.  Color bars show velocity and
brightness scales in each image.

\noindent
{\bf Figure~5.~~} {\cof} channel maps at 1.3\kms\ spectral resolution.
The central velocity is indicated in the upper left of each panel.
The channel RMS is 60\mjyb, the peak is 3.3\jyb, and contours are
plotted at $\pm 5, 15, 20, 30, 40$ and $50\, \sigma$.  The lower right
panel shows the synthesized beam in the bottom right corner ($3.83''
\times 3.45''$ at P.A. $-59.7^\circ$) and a scale size of 0.55 pc.
Plus symbols represent the locations of the UC\,H{\sc ii} region at
the 6~cm continuum peak and G34.4~MM as indicated in the upper left
panel.

\noindent
{\bf Figure~6.~~} {\tcof} channel maps at 1.36\kms\ spectral
resolution.  The central velocity is indicated in the upper left of
each panel.  The channel RMS is 44\mjyb, the peak is 1.7\jyb, and
contours are plotted at $\pm 3, 5, 7, 9, 12, 15, 20$ and $25\,
\sigma$.  The lower right panel shows the synthesized beam in the
bottom right corner ($4.04'' \times 3.83''$ at P.A. $-49.9^\circ$) and
a scale size of 0.55 pc.  Plus symbols represent the locations of the
UC\,H{\sc ii} region at the 6~cm continuum peak and G34.4~MM as
indicated in the upper left panel.

\noindent
{\bf Figure~7.~~} {\ceof} channel maps at 0.7\kms\ spectral
resolution.  The central velocity is indicated in the upper left of
each panel.  The channel RMS is 60\mjyb, the peak is 0.92\jyb, and
contours are plotted at $\pm 3, 5, 7, 9, 11, 13$ and $15\, \sigma$.
The lower right panel shows the synthesized beam in the bottom right
corner ($4.04'' \times 3.83''$ at P.A. $-50.1^\circ$) and a scale size
of 0.55 pc.  Plus symbols represent the locations of the UC\,H{\sc ii}
region at the 6~cm continuum peak and G34.4~MM as indicated in the
upper left panel.

\noindent
{\bf Figure~8.~~} Spectra convolved with a $10''$ beam taken in CO,
{\tco} and {\ceo} at different locations in the image.  The ratio of
the CO and {\tco} emission was used to calculate optical depth in the
red and blue-shifted outflow lobes as indicated by the dashed-dot
lines and arrows in the figure.  Panels i) \& ii) show spectra taken
near the G34.4 UC\,H{\sc ii} region at positions \ra{18}{53}{18.320}\,
\dec{1}{24}{45.84} and \ra{18}{53}{19.203}\, \dec{1}{24}{39.38},
respectively.  Panels iii) \& iv) show spectra taken near the G34\,MM
core at positions \ra{18}{53}{18.065}\, \dec{1}{25}{25.06} and
\ra{18}{53}{17.284}\, \dec{1}{25}{13.52}, respectively.  The locations
of the spectra relative to the {\tco} emission is shown in the center
panel of Figure 4.  The dashed vertical line at 57{\kms} helps to
illustrate that the peak of the {\ceo} line shifts from 57.4{\kms}
toward the G34\,MM core to 56{\kms} toward the G34.4 UC\,H{\sc ii}
region core.

\noindent
{\bf Figure 9.}~~GLIMPSE 4.5\,$\mu$m image in inverted greyscale
showing locations of G34.4\,MM and G34.4 UC\,H{\sc ii} region as
crosses.  Open diamonds show the predicted locations of the driving
sources of outflows G34.4:D and E based on CO and {\tco} morphology.
GLIMPSE sources are located very near the predicted positions
suggesting that GLIMPSE has detected two sources driving outflows D
and E.  {\tco} red- and blue-shifted emission contours from Fig. 4 are
overlaid for reference.

\noindent
{\bf Figure 10.}~~GLIMPSE and MIPSGAL 3-color image with 4.5\,$\mu$m
(blue), 8.0\,$\mu$m (green) and 24\,$\mu$m (red).  The probable YSO
cluster members listed in Table 3 are marked by color-coded circles:
green = good YSO fits ($\chi^2/N_{\rm data} \leq 4.8$), yellow = bad
YSO fits ($\chi^2/N_{\rm data} > 14.8$), cyan = source detected in
fewer than four bands, and green = MIPS 24\,$\mu$m point sources
detected in too few IRAC+2MASS bands to be fit. Sources well-fit by
stellar photospheres are marked by blue circles. The IR dark cloud has
been divided into 3 regions for the purpose of the YSO census: Region
C (large white circle) contains the OVRO observations and regions N
and S are (large white rectangles) are the portions of the dark cloud
lying to the north and south, respectively, of region C. Source
numbers from Table 3 for the YSOs discussed in Section 3.4 are also
shown. 

\noindent
{\bf Figure 11:} (a) Ten best YSO SED model fits to the IR fluxes of
source 12/G34.4\,MM.  The fitting assumes that the mid-IR flux from
the millimeter core is dominated by a single source. The [4.5] upper
limit (inverted triangle) was used because the source appears to have
excess flux in this bandpass due to molecular emission associated with
outflow activity that has not been incorporated into the YSO
models. The thickest line represents the best fit to the data, and
different colors represent the SED observed through different
apertures, from smallest (IRAC, black) to largest ({\it MSX} 21.3 \um,
green). The dashed line shows the photospheric flux from the best-fit
YSO model, as it would appear if suffering from interstellar
extinction only in the absence of circumstellar material.  (b) Masses
and luminosities of the best fit models.  The shaded contour regions
represent the logarithmic density of models in the mass-luminosity
parameter space. The darkest gray represents a model density 1000
times higher than that of the lightest grey. The white region beyond
the lowest contour shows the part of parameter space that lies beyond
the edge of the model grid.

\noindent
{\bf Figure 12:} (a) Same as Figure 11a, but for the other 6 well-fit
YSOs in region C.  All fits used to compute the model parameter ranges
in Table 4 are plotted. Sources 10 and 11 overlap the \uchii\ region,
and hence the diffuse mid-IR fluxes associated with the \uchii\ region
in all 4 {\it MSX} bands and MIPS 24 and 70 \um\ were used as upper
limits to the fits. (b) Same as Figure 11b, but for the 6 well-fit
YSOs plotted in Figure 12a.

\newcommand{\gname}[2]{\includegraphics[width=#2in]{#1.jpg}}

\clearpage
\begin{figure}
\gname{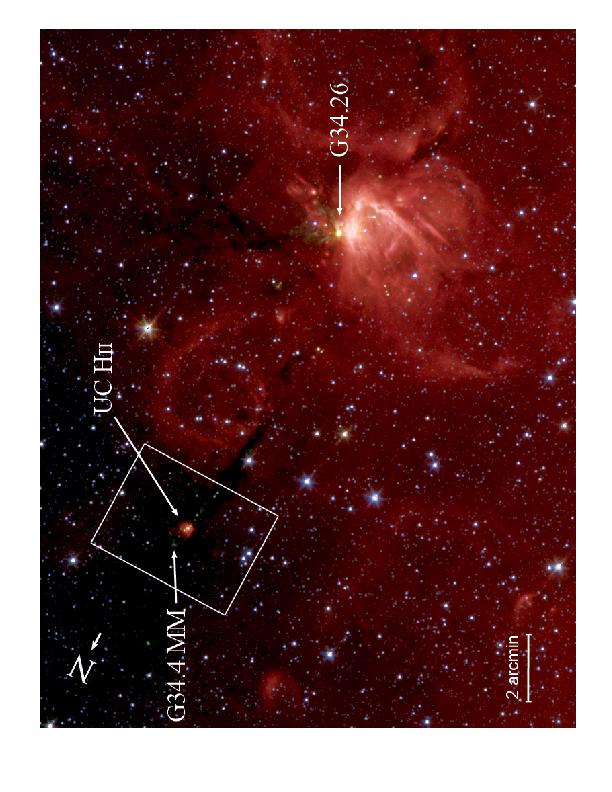}{6}
\center{\bf Figure 1.}
\end{figure}

\clearpage
\begin{figure}
\gname{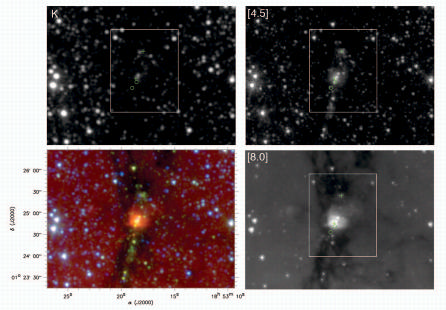}{6}
\center{\bf Figure 2.}
\end{figure}

\clearpage
\begin{figure}
\gname{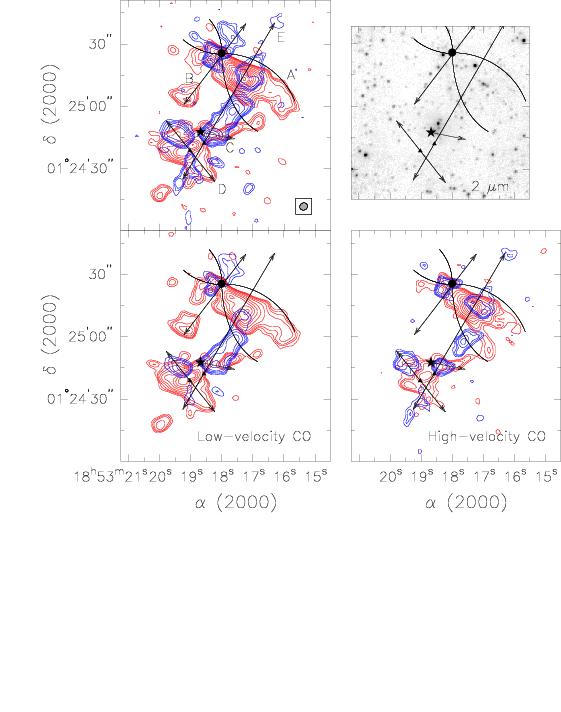}{6}
\center{\bf Figure 3.}
\end{figure}

\clearpage
\begin{figure}
\gname{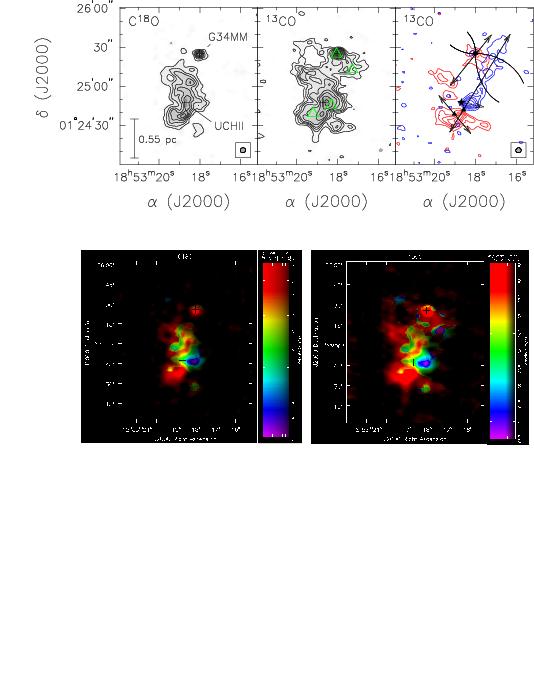}{6}
\center{\bf Figure 4.}
\end{figure}

\clearpage
\begin{figure}
\gname{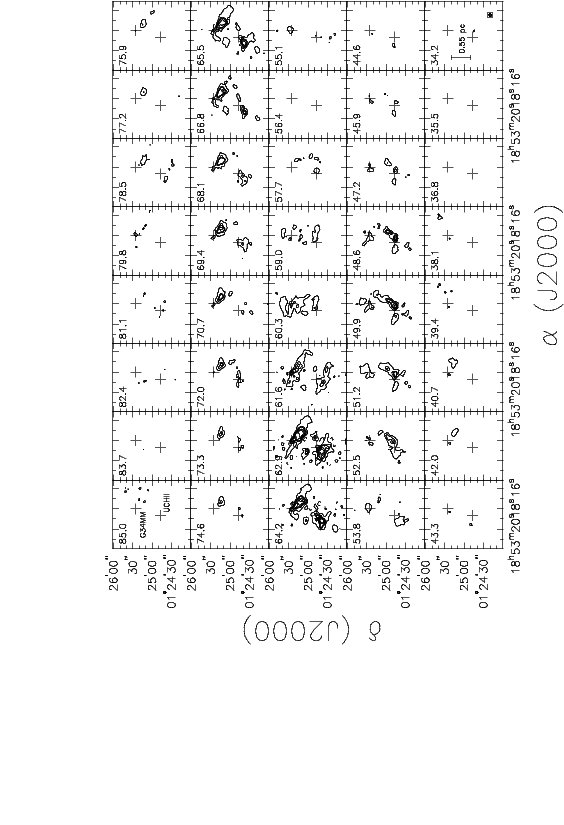}{6}
\center{\bf Figure 5.}
\end{figure}

\clearpage
\begin{figure}
\gname{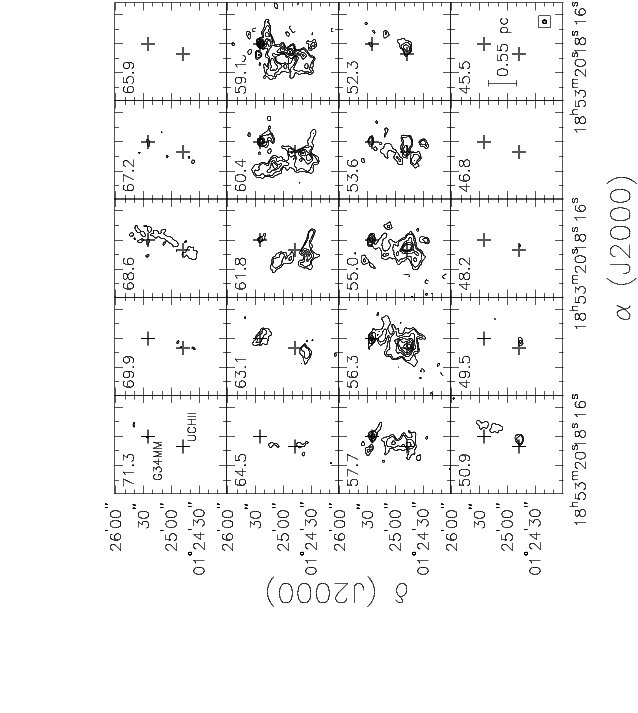}{6}
\center{\bf Figure 6.}
\end{figure}

\clearpage
\begin{figure}
\gname{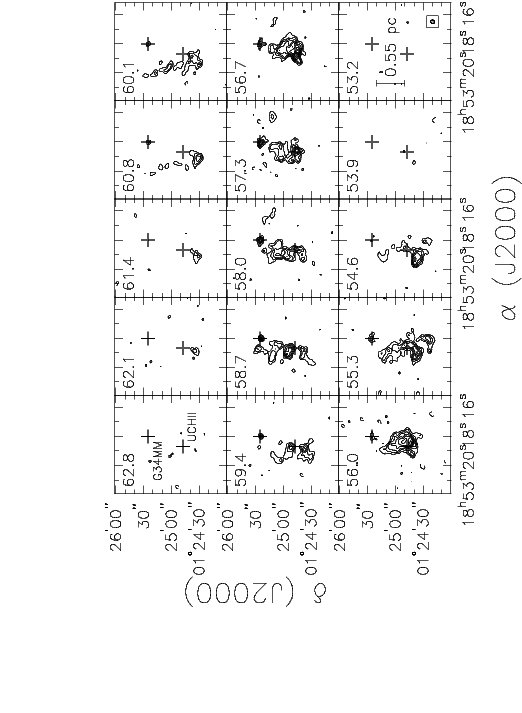}{6}
\center{\bf Figure 7.}
\end{figure}

\clearpage
\begin{figure}
\gname{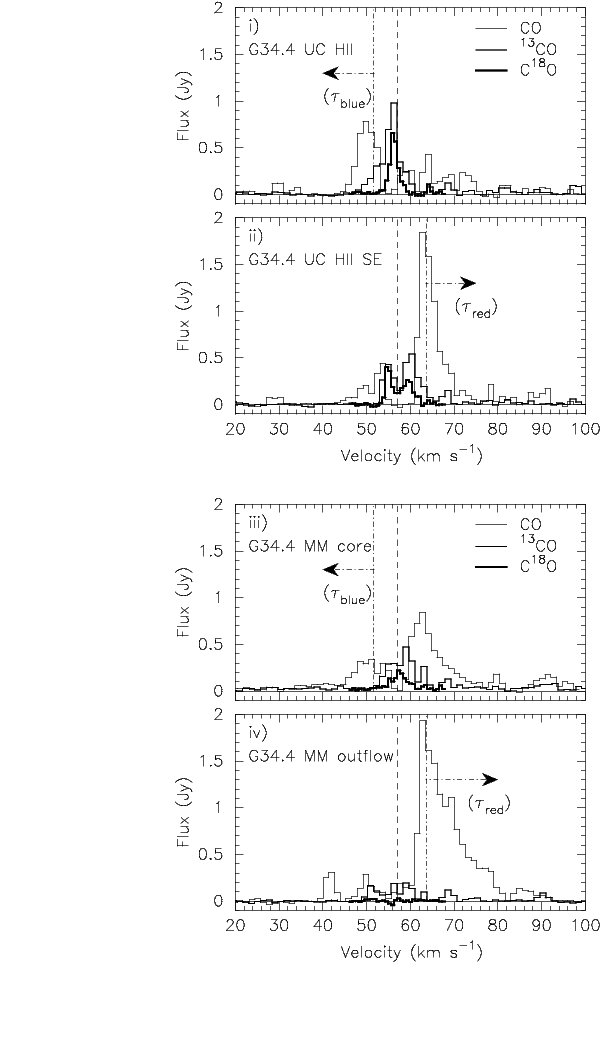}{5}
\center{\bf Figure 8.}
\end{figure}


\clearpage
\begin{figure}
\gname{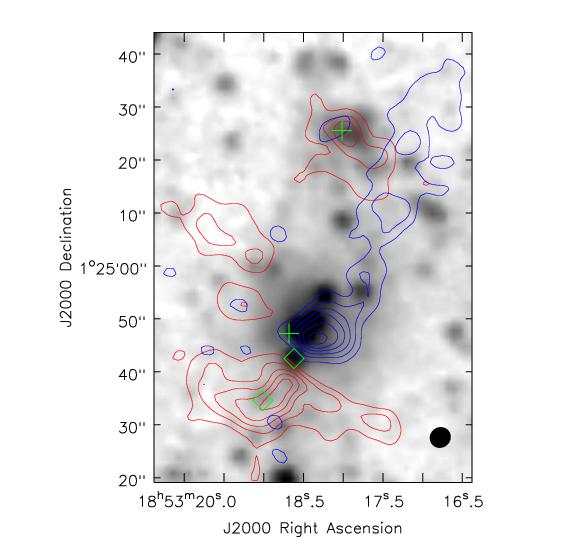}{6}
\center{\bf Figure 9.}
\end{figure}


\clearpage
\begin{figure}
\gname{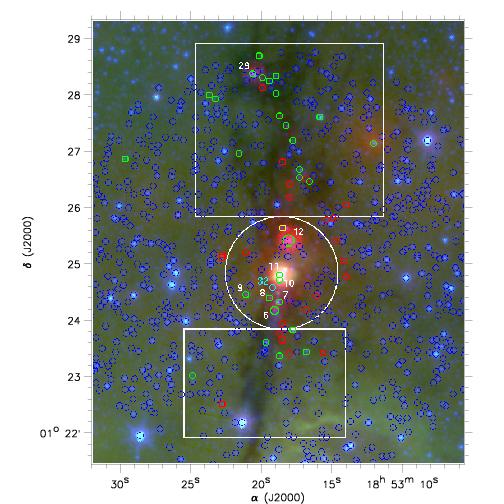}{6}
\center{\bf Figure 10.}
\end{figure}

\clearpage
\begin{figure}
\gname{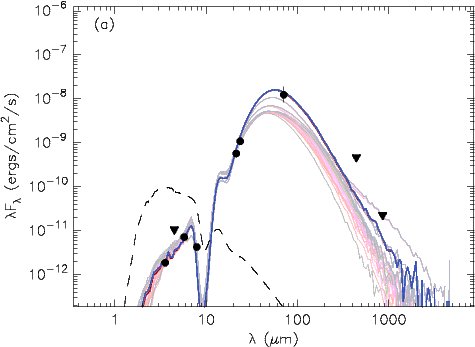}{6}
\center{\bf Figure 11a.}
\end{figure}

\clearpage
\begin{figure}
\gname{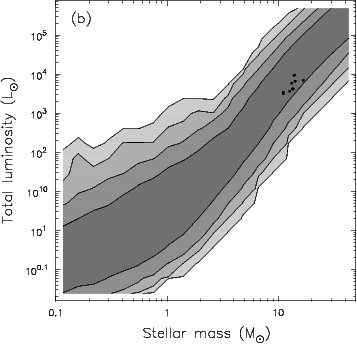}{6}
\center{\bf Figure 11b.}
\end{figure}

\clearpage
\begin{figure}
\gname{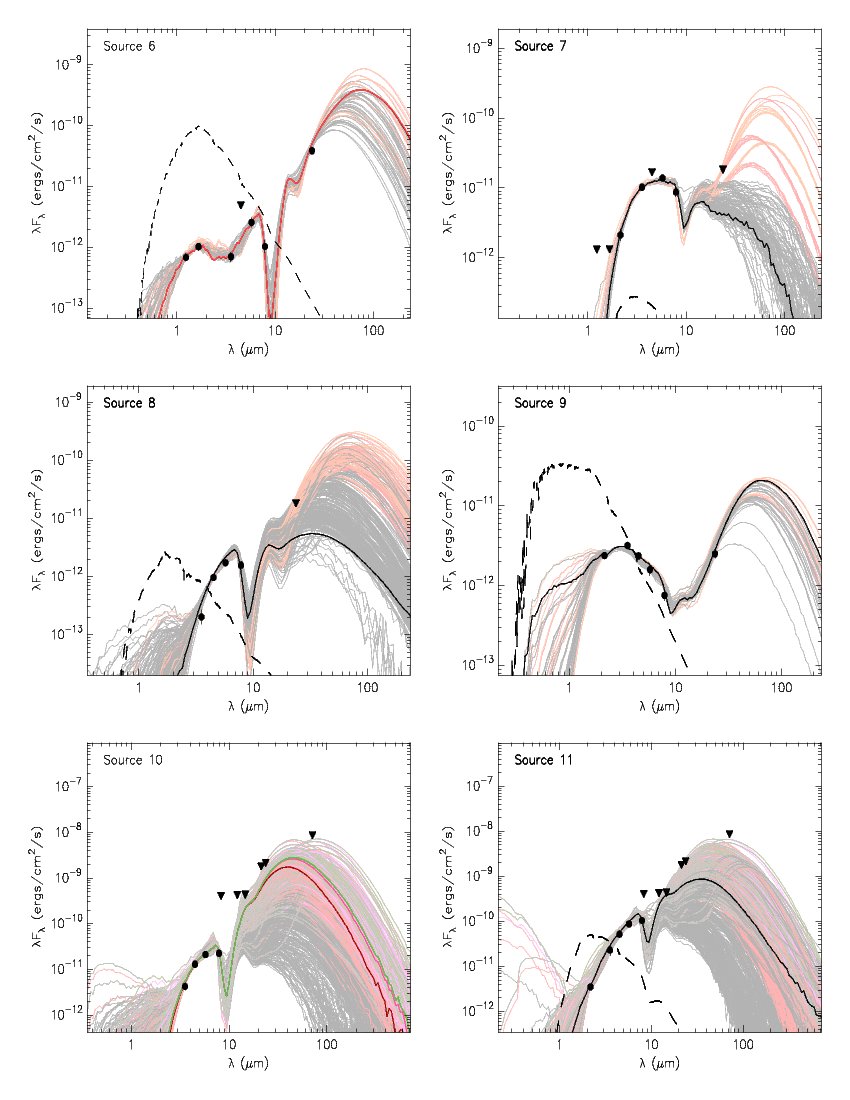}{6}
\center{\bf Figure 12a.}
\end{figure}

\clearpage
\begin{figure}
\gname{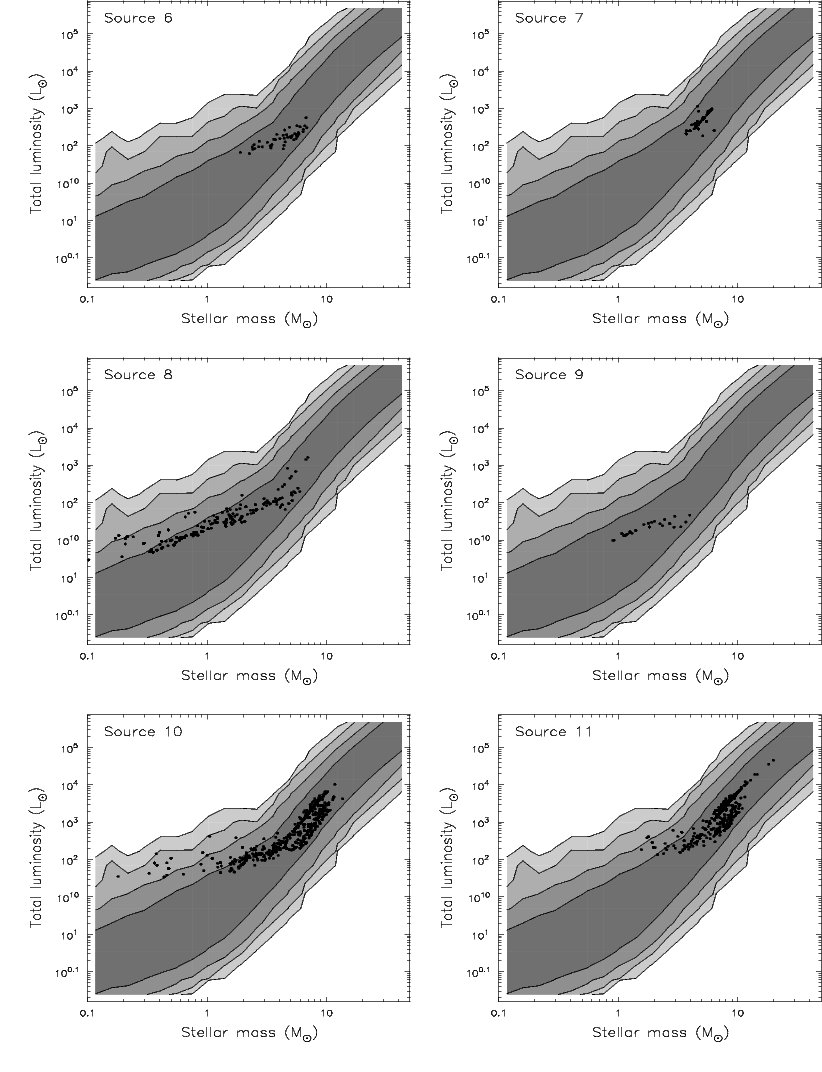}{6}
\center{\bf Figure 12b.}
\end{figure}

\end{document}